\documentclass[showpacs,prb,letterpaper,floatfix,nobalancelastpage,twocolumn]{revtex4}
\usepackage{dcolumn}
\usepackage{bm}
\usepackage{amsmath}
\usepackage{bm,braket}
\usepackage{color}
\usepackage{float}
\usepackage{graphicx,subfigure}
\usepackage{amssymb}
\usepackage{sidecap}

\begin{document}


\title{Theory of the quantum-dot Mollow triplet in an exciton-driven semiconductor cavity}

\author{C. Roy}
\email{chiranjeeb.roy@queensu.ca}
\author{S. Hughes}
\email{shughes@physics.queensu.ca}
\affiliation{Department of Physics, Engineering Physics and Astronomy,
Queen's University, Kingston, Ontario, Canada K7L 3N6}

\begin{abstract}

We present a comprehensive theoretical study of the resonance fluorescence spectra of an exciton-driven quantum dot (QD) placed inside a high-$Q$ semiconductor cavity and interacting with an acoustic phonon bath.  We derive a quantum master equation (ME) in the polaron frame which includes exciton-phonon and exciton-cavity coupling to all orders.
This work details and extends the  theory
used in a recent issue of {\em Physical Review Letters} (C. Roy and S. Hughes
2011: Phys.~Rev.~Lett. {\bf 106}
247403) to describe
the QD Mollow triplet in the regime of semiconductor cavity-QED.
 Here we introduce two ME forms, Nakajima-Zwanzig (NZ) and time-convolutionless (TC), both to second order in the system--phonon-reservoir  perturbation. In the polaron frame, these two ME forms are shown to yield equivalent population dynamics and fluorescence spectra for a continuous wave (cw) driving field. We also demonstrate that a Markov approximation is valid for computing the incoherent  scattering processes and we subsequently exploit the  Markovian TC ME to explore the resonance fluorescence spectra of an exciton-driven QD. Both cavity-emitted and exciton-emitted spectra are studied  and these are found to have qualitatively different spectral features.
 Using a  coherent driving  field, the well known characteristics of the atomic Mollow triplet are shown to be considerably modified with electron--acoustic-phonon scattering and
we highlight the key effects arising from both cavity coupling and electron-phonon coupling.
 Regimes of pronounced cavity feeding and anharmonic cavity-QED are exemplified, and we find that the cavity coupling depends sensitively on the exciton-cavity detuning and the temperature of the phonon bath. We show how the full width at half maximum  (linewidth) of the Mollow triplet sidebands varies as a function of  the square of the Rabi frequency of the cw pump. Phonon-mediated cavity coupling also contributes to the spectral broadening of the Mollow triplet, depending upon the exciton-cavity detuning and the strength of the exciton-cavity coupling rate. Finally, we calculate the fluorescence spectra for off-resonance cw driving and investigate the resulting  Mollow-triplet linewidths.
\end{abstract}


\maketitle

\section{Introduction}

Semiconductor quantum dots (QDs) interacting with microcavities are a field of intense research where ideas from conventional cavity-quantum electrodynamics (cavity-QED) are being applied and studied~\cite{yamamoto,hohenester}. These chip-based  semiconductor devices show enormous promise for technological applications ranging from quantum information processing and quantum metrology to the creation of single photon on demand \cite{press,entangled1,entangled2,entangled3,SC:StrongCoupling1,SC:StrongCoupling2,SC:StrongCoupling3}. The possibility of exploiting quantum {\em anharmonicities}  in these cavity-QED  systems may also lead to photon blockade~\cite{spinblockade}. Additionally, the spin degree of freedom of the underlying exciton states is of interest for quantum information processing \cite{atature}. In the past few years, an increasing number of  experimental studies have focussed on resonance fluorescence of a QD coupled to a cavity mode~\cite{muller,flagg,vamivakas,ates1}. Progress has  been made in the study of off-resonant QD cavity system that can be used to observe the photoluminescence (PL) spectra of the QD or the cavity~\cite{ates2,jelena1,ates3}.

Recent experimental studies of resonance fluorescence of InGaAs QDs embedded in a high-quality micropillar cavity~\cite{stuttgart_prl} have provided clear demonstration of excitation-induced dephasing (EID).
In QD structures with no cavity coupling, the 
phonon-induced EID process is responsible for the intensity damping of the QD excitonic Rabi rotations induced by pulsed excitation~\cite{ramsay1,ramsay2}. The experimental resonance fluorescence spectra in QD-cavities is similar to the characteristic Mollow triplet of an atomic system, but the sidebands of the triplet are found to undergo systematic spectral sideband broadening with increasing continuous-wave (cw) drives \cite{stuttgart_prl}. This drive-dependent broadening  results from the interaction of the driven QD with the underlying phonon reservoir~\cite{besombes,Favero:PRB03,Peter:PRB04,nazir2,mogilevtsev1,mogilevtsev2}. For cavity-QED (quantum electrodynamics) systems, acoustic phonon scattering processes also introduce an additional coupling mechanism between a cavity mode and the exciton~\cite{roy_hughes,jelena_arka}. This phonon-mediated cavity coupling depends on the QD-cavity detuning and the intrinsic parameters of the phonon bath. Cavity-exciton coupling can be used to measure the fluorescence spectra of the QD via the cavity mode which is convenient because of spectral (and possible geometrical) separation between the exciton and the cavity mode. 

Acoustic phonon processes also play an important role in {\em incoherently} pumped cavity-QED systems which are known to result in off-resonant ``cavity feeding''~\cite{HennessyNature:2007,KaniberPRB:2008,RuthOE:2007,SufczynskiPRL:2009,ota,TawaraOE:2009,Dalacu:PRB2010,Calic:PRL11}, an effect which  also exists (and is particularly strong) in coherently driven systems~\cite{roy_hughes,jelena_arka}. There have been several attempts at an improved theoretical description of a semiconductor cavity-QED system to include the effects of electron--acosutic-phonon scattering~\cite{JiaoJPC2008,ota,HohenesterPRB:2009,HohenesterPRB:2010,kaer,hughes2,SavonaPRB:2010}. As an example, the vacuum Rabi doublet of a cavity-QED system in the strong coupling regime  undergoes modification due to phonon coupling~\cite{hughes1}, e.g., the  spectral doublets---for on-resonance coupling---are {\em asymmetric}, especially at low temperatures, which can be explained in terms of phonon-assisted processes~\cite{imamoglu,ota,kaer,hughes2}.

The complexities of modelling a coherently-driven semiconductor cavity-QED system (where multiphoton effects are required) are best treated in a quantum master equation (ME) formalism which can be used to investigate both qualitative and quantitative changes in the Mollow spectra due to phonon-induced scattering. In this paper, we
present the theoretical details behind the model
used in a recent edition of {\em Physical Review Letters}~\cite{roy_hughes}. We then
 use this model to systematically explore the exciton-driven 
 Mollow triplet in  a wide range of 
 excitation regimes.
 We first present two alternative ME forms which govern the evolution of the reduced density operator, wherein we include the system-bath incoherent interaction to second order~\cite{roy_hughes,nazir2,me_book_1} in the system-reservoir coupling. These two MEs  are termed the Nakajima-Zwanzig (NZ) form and the time-convolutionless (TC) form. To second-order in the polaron transformed
 system-phonon-reservoir coupling, these ME forms are equivalent to the standard Born and Born-Markov MEs, but we keep the former notation (TC) to avoid confusion with a second-Markov approximation which can be applied on the memory relaxation
 of the phonon reservoir.  The polaron frame, obtained following a unitary polaron-transformation, allows us to study the effects of phonon dephasing on the coherent part of the Hamiltonian to all orders, i.e., beyond the Born approximation. Moreover, these polaron MEs allow one to study the nonperturbative dynamics of a cavity-QED system interacting with a radiation reservoir and driven by an external cw laser field which is also under the influence of the phonon environment. The use of the polaron frame eliminates the QD-phonon coupling in the system Hamiltonian and introduces a modified exciton-cavity coupling and a modified radiative decay rate~\cite{mahan,krum,imamoglu}. 

The rest of our paper is organized as follows. In Sec.~\ref{model} we present the model Hamiltonian and derive a second-order TC   form of the ME and provide a  comparison with the second-order NZ  form. The computational effort necessary to solve the NZ ME is significantly more complicated than the TC ME. However, for cw drives, we show that both ME forms produce the same system dynamics and spectra. 
We also study the  Markov limit of the TC ME by neglecting the memory dynamics of the phonon reservoir, and show that, for our parameters and drives,  the full non-Markovian solution is identical (or at least indistinguishable) to the dynamics obtained in the Markov limit. We subsequently use the Markov form of the TC ME form for our remaining calculations. In Sec.~\ref{results} we present and discuss our numerical results of the resonance fluorescence spectra of a  coherently driven cavity-QED system with an exciton pump field. The QD linear absorption spectrum is first shown for several different bath temperatures. We then study the role of the temperature of the underlying phonon reservoir on the fluorescence spectrum. We  explore the effects of QD-cavity detuning and identify regions where the QD-cavity coupling is particularly strong. Next, we provide an analysis of the effects of the cw-laser drive and QD-cavity coupling on the spectrum and  present detailed simulations of the fluorescence spectra for various QD-cw laser detuning. We subsequently investigate EID effects by numerically extracting the spectral linewidths of the Mollow triplet sidebands for various cw-laser drives. In addition, we study the dependence of QD-laser detuning on the Mollow triplet broadening. In Sec.~\ref{conclusions}, we summarise our results and present our conclusions.

\section{Theory}
\label{model}
\subsection{Model Hamiltonian}

We start by introducing a model Hamiltonian which describes the exciton-driven cavity-QED system, where the QD interacts with a phonon reservoir and the dot is driven by a classical cw-laser field. Considering a single  exciton (electron-hole pair), and transforming to a frame rotating with respect to the laser pump frequency, $\omega_L$, the  model Hamiltonian is  \begin{eqnarray}
\label{sec1eq1}
H&=&\hbar\Delta_{xL}{\sigma}^{+}{\sigma}^{-}+\hbar\Delta_{cL}{a}^{\dagger}{a}\nonumber \\
& & +\hbar g({\sigma}^{+}{a}+{a}^{\dagger}{\sigma}^{-}) + \hbar \eta_{x}({\sigma}^{-}+{\sigma}^{+})\nonumber \\
& & +{\sigma}^{+}{\sigma}^{-}\sum_{q}\hbar\lambda_{q}({b}_{q}
+{b}_{q}^{\dagger})+\sum_{q}\hbar\omega_{q}{b}_{q}^{\dagger}{b}_{q},
\end{eqnarray}
where the phonon bath is represented by a continuum of harmonic oscillators
of frequency $\omega_q$ with annihilation (creation) operators
${b}_{q}({b}_{q}^{\dagger})$, and $\lambda_q$ are  the exciton-phonon couplings (assumed real); also introduced in Eq.~(\ref{sec1eq1}) are the 
detunings of 
the exciton ($\omega_{x}$) and cavity ($\omega_{c}$) from the  pump laser frequency,
 $\Delta_{\alpha L}\equiv \omega_\alpha-\omega_L$ ($\alpha =x,c$), 
 the  cavity mode annihilation (creation) operator, $a(a^{\dagger})$,
 the Pauli operators of the  exciton, 
 ${\sigma}^+,{\sigma}^-$, and  the exciton-cavity  coupling rate,
  $g$. The exciton pumping rate, $\eta_{x}$, describes the coherent exciton pumping from the cw laser field. We have included the role of phonons at the level of the independent boson model (IBM) ~\cite{mahan}, which is known to yield good agreement with experiments on solid-state QDs~\cite{hohenester}. A more complete description of the phonon reservoir, e.g., to allow broadening of the zero phonon line (ZPL), would require the generalization of the above system Hamiltonian to include additional microscopic processes, such as spectral diffusion, anharmonicity effects \cite{10,11}, phonon scattering from interfaces \cite{12,13}, and a modified phonon spectrum \cite{14}.
Phenomenologically, we  include radiative and pure dephasing mechanisms to broaden the ZPL, while radiative coupling to the cavity mode is treated in a self-consistent way.


\begin{figure}[t!]
\centering\includegraphics[width=.99\columnwidth]{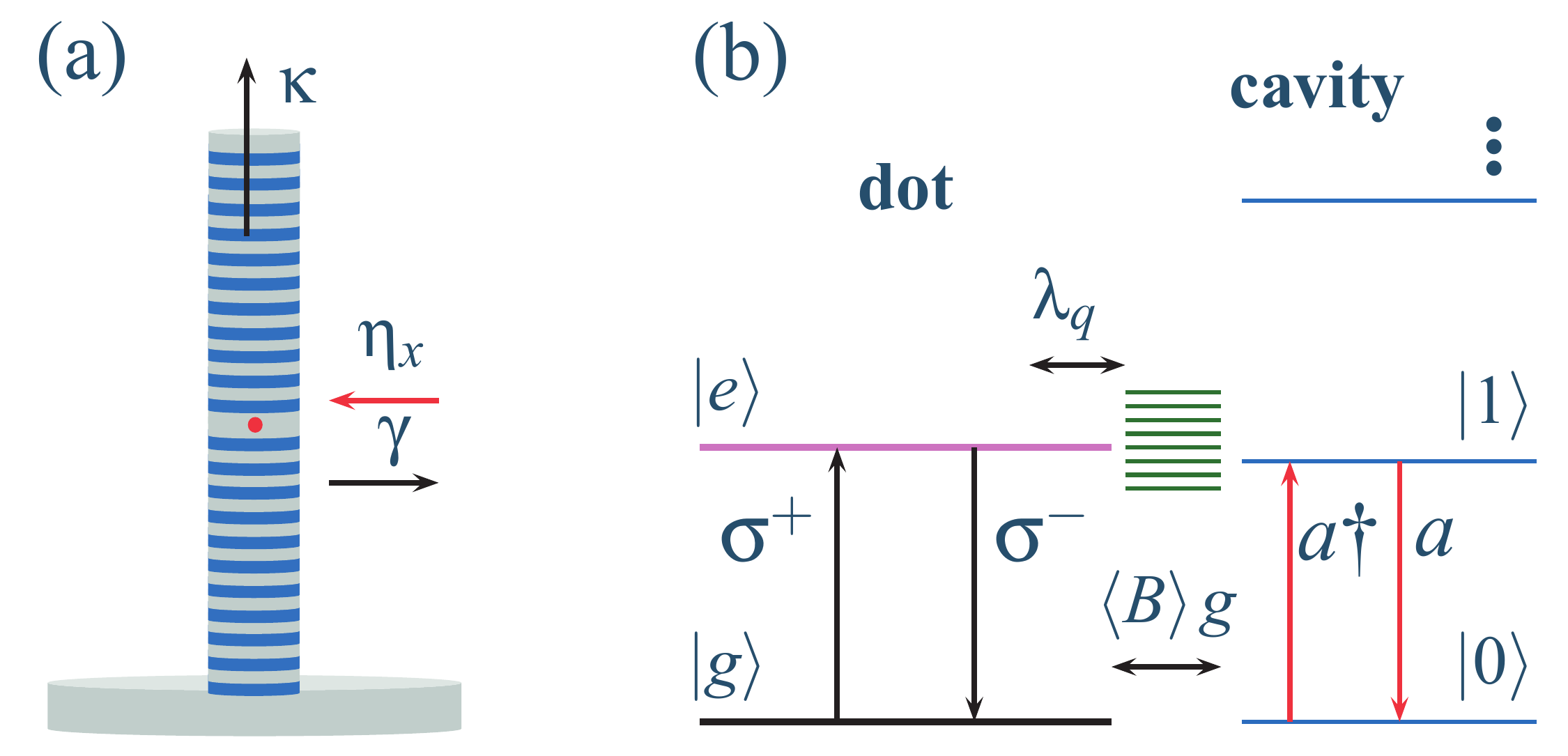}
\label{fig:schematic}
\caption{(Color online)
(a) Schematic of a semiconductor cavity used in cavity-QED (micropillar system), containing a coupled QD and driven by a coherent cw laser ($\eta_x$ from the side). The fluorescence spectra of this QD-driven system is measured via cavity emission ($\kappa$) or QD emission ($\gamma$).  (b) Energy\\ level diagram, where $\vert e\rangle$ denotes the excited QD state, $\vert g\rangle$ denotes the ground QD state, and $\ket{0,1,\cdots}$ show the cavity photon states. Also shown is the phonon reservoir (e.g., a bath of harmonic oscillator states treated at the level of the IBM) as green  lines which is  coupled to the QD excited state.}
\end{figure}

We seek to solve the Hamiltonian dynamics in the polaron frame~\cite{hohenester}, where $H' \rightarrow e^{P} H e^{-P}$, with $P= \sigma^+ \sigma^-\sum_q \frac{\lambda_q}{\omega_q} ( b_q^\dagger- b_q).$ 
Closely following the notation of Wilson-Rae and Imamo\v glu~\cite{imamoglu}, the polaron-transformed Hamiltonian can be written in terms of a modified system, bath, and interaction part:
\begin{subequations}
\begin{align}
\label{sec3eq1}
H^{\prime}_{S}  & =    \hbar(\Delta_{xL}-\Delta_{P}){\sigma}^{+}{\sigma}^{-}
+\hbar\Delta_{cL} {a}^{\dagger}{a}+\langle B\rangle {X}_{g}, \\
H^{\prime}_{B} & =  \sum_{q}\hbar\omega_{q}{b}_{q}^{\dagger}{b}_{q},\\
H^{\prime}_{I}  & =  {X}_{g}{\zeta}_{g}+{X}_{u}{\zeta}_{u},
\end{align}
\end{subequations}
where $ B_{\pm}$ are the coherent displacement operators of the phonon modes,
\begin{align}
\label{sec3eq2a}
{B}_{\pm}&=\exp\left [\pm\sum_{q}\frac{\lambda_{q}}{\omega_{q}}\left ({b}_{q}-{b}_{q}^{\dagger}\right) \right],
\end{align}
\\
and the  fluctuation operators are defined through,
$
{\zeta}_{g}=\frac{1}{2}({B}_{+}+{B}_{-}-2\langle B\rangle)$ and ${\zeta}_{u}=\frac{1}{2i}({B}_{+}-{B}_{-})$.
The thermally-averaged phonon displacement operators  obey the relation~\cite{mahan} $\langle B\rangle=\langle  {B}_{+}\rangle=\langle { B}_{-}\rangle$, 
where
\begin{align}
\label{sec3eq5a}
\langle  B\rangle&=\exp\left[-\frac{1}{2}\int^{\infty}_{0}d\omega\frac{J(\omega)}
{\omega^{2}}\coth(\beta\hbar\omega/2)\right ], 
\end{align}
and $J(\omega)=\sum_q \lambda_q^2\delta(\omega-\omega_q)$ is the phonon spectral function that
will be  explicitly defined later (in a more practical continuum form).
 The polaron frequency shift, given by $\Delta_{P}=\sum_{q}{\lambda_{q}^{2}}/{\omega_{q}}
=\int^{\infty}_{0}d\omega{J(\omega)}/{\omega}$, accounts for the renormalization of the QD exciton resonant frequency due to coupling to the phonon modes; however, we will assume that the polaron shift is implicitly included in $\omega_{x}$.
The phonon-bath-modified system operators, $ X_{g}$ and $ X_{u}$, are given by
\begin{subequations}
\begin{align}
\label{sec3eq2b}
{X}_{g}& = \hbar g({a}^{\dagger}{\sigma}^{-}+{\sigma}^{+}{a})
+\hbar \eta_{x}({\sigma}^{-}+{\sigma}^{+}),  \\
{X}_{u}& = i\hbar g({\sigma}^{+}{a}-{a}^{\dagger}{\sigma}^{-})
+i\hbar \eta_{x}({\sigma}^{+}-{\sigma}^{-}),
\end{align}
\end{subequations}
 which results in a  slightly different definition of the system Hamiltonian given in Eq.~(\ref{sec3eq1}); this modified system Hamiltonian is central to obtaining results in the nonperturbative phonon-coupling regime~\cite{carmichael,imamoglu,roy_hughes_prx}.

Figure 1 shows a schematic of an example QD-cavity system [Fig.~1(a)], as well as an energy
level diagram  showing the exciton levels [Fig.~1(b)], some of the phonon levels, and the  cavity photon states.
The effective exciton-cavity coupling term, $\braket{B}g$, corresponds to a coherent reduction of $g$ due to acoustic phonon interactions [cf.~Eq.~(\ref{sec3eq1})].

\subsection{Master equation}

%

The usual quantum optics approach to describe system-reservoir interactions is to derive a ME in a form suitable for numerical (or analytical) solution  that can include photon interactions to any arbitrary order. The ME is usually derived using projection operator techniques such as the NZ approach or the TC approach~\cite{me_book_1}. The former technique leads to an integro-differential ME convoluted with the memory kernel which encapsulates the effects of the reservoir. However, the latter technique (TC) results in a ME which is \textit{local} in time with time-dependent superoperators acting on the reduced density operator~\cite{me_1,me_2,me_3}. 

The derivation of the ME for the reduced density operator of the system in the NZ form starts with an equation that is written to second-order in the system-reservoir couplings (Born approximation)~\cite{me_4} in the interaction picture. Defining the interaction Hamiltonian,  
\begin{align}
\tilde{H}_{I}(t)=e^{i(H_{S}^{\prime}
+H_{B}^{\prime})t}H_{I}'e^{-i(H_{S}^{\prime}+H_{B}^{\prime})t},
\label{eq:HI}
\end{align} 
then we obtain the following equation of motion for the reduced density operator,
\begin{align}
\label{sec_new_eq1}
\left .\frac{\partial \tilde{\rho}(t)}{\partial t}\right |_{\rm NZ}=-\frac{1}{\hbar^{2}}\int^{t}_{0}dt^{\prime}\,\rm{tr}_{B}
\left \{\left [\tilde{H}_{I}(t),\left [\tilde{H}_{I}(t^{\prime}), \tilde{\rho}(t^{\prime})\rho_{B}\right ] \right ]\right \},
\end{align}
where the tildes denote the operators in the interaction picture and `tr' is the trace. A Markovian form of Eq.~(\ref{sec_new_eq1}), to second order  in the interaction terms, can be derived by approximating $\tilde{\rho}(t^{\prime})$ with $\tilde{\rho}(t)$, so that
\begin{align}
\label{sec3eq3}
\left . \frac{\partial \tilde{\rho}(t)}{\partial t} \right |_{\rm TC}=
-\frac{1}{\hbar^{2}}\int^{t}_{0}dt^{\prime}\,\rm{tr}_{B}\left \{\left [\tilde{H}_{I}(t),\left [\tilde{H}_{I}(t^{\prime}), \tilde{\rho}(t)\rho_{B}\right ] \right ]\right \}.
\end{align}
This TC form eliminates the $\tilde\rho(t')$ dependence of the integral on the right-hand side of the density operator equation at all previous times~\cite{carmichael,me_book_1}. 
 However, some memory (time history) effects can still exist depending upon the form of the bath interaction. A further simplification to Eq.~(\ref{sec3eq3}) is obtained by extending the upper limit of the integral to $\infty$. This is a \textit{second-Markov approximation} that yields a ME in the TC Markov form. In fact, this latter (simpler) ME form can be   further manipulated---under certain approximations---to produce an approximate phonon ME of the standard Lindblad form~\cite{roy_hughes_prx}; we have previously demonstrated such
an approach to study intensity power broadening.


Substituting the interaction Hamiltonian [Eq.~\ref{eq:HI}] 
into  Eq.~(\ref{sec3eq3}), gives
\begin{align}
\label{sec3eq5}
\left . \frac{\partial \tilde{\rho}(t)}{\partial t} \right |_{\rm TC}&=-\frac{1}{\hbar^{2}}\sum_{m=g,u}\nonumber \\
& \left [
\int^{t}_{0}dt^{\prime}{\rm tr}_B\left \{\zeta_{m}(t)\zeta_{m}(t^{\prime})\rho_{B}\right \}\tilde{X}_{m}(t)\tilde{X}_{m}(t^{\prime})\tilde{\rho}(t)
\right .
\nonumber \\
&-\int^{t}_{0}dt^{\prime}{\rm tr}_B\{\zeta_{m}(t)\rho_{B}\zeta_{m}(t^{\prime})\}\tilde{X}_{m}(t)\tilde{\rho}(t)\tilde{X}_{m}(t^{\prime}) \nonumber \\
&-\int^{t}_{0}dt^{\prime}{\rm tr}_B\{\zeta_{m}(t^{\prime})\rho_{B}\zeta_{m}(t)\}\tilde{X}_{m}(t^{\prime})\tilde{\rho}(t)\tilde{X}_{m}(t)
\nonumber \\
&
\left . +\int^{t}_{0}dt^{\prime}{\rm tr}_B\{\rho_{B}\zeta_{m}(t^{\prime})\zeta_{m}(t)\}
\tilde{\rho}(t)\tilde{X}_{m}(t^{\prime})\tilde{X}_{m}(t) \right ],
\end{align}
where $\zeta_{m}(t)=e^{iH_B^{\prime}t}\zeta_{m}e^{-iH_B^{\prime}t}$. The separation of the trace over the phonon variables is possible because we assume  an initially separable density operator consisting of the system (reduced) density operator and the phonon bath density operator $\rho_{B}$.  We have also assumed that the full density operator [i.e, $\rho_B\rho(t)$] remains separable at all later times. We denote ${\rm tr}_B\{\rho_{B}\zeta_{m}(t)\zeta_{m}(t^{\prime})\}=G_{m}(t-t^{\prime})$ and ${\rm tr}_B\{\rho_{B}\zeta_{m}(t^{\prime})\zeta_{m}(t)\}=G_{m}(t^{\prime}-t)$, where  $G_{m}(t)$ are the  polaron Green functions~\cite{mahan,imamoglu}:
\begin{subequations}
\begin{align}
G_{g}(t)&=\langle  B\rangle^{2}\left (\cosh[\phi(t)]-1\right ),\\ 
G_{u}(t)&=\langle  B\rangle^{2}\sinh[\phi(t)], 
\label{sec3eq4}
\end{align}
\end{subequations}
and the phonon correlation function $\phi(t)$ is defined as
\begin{align}
\label{sec3eq5b}
\phi(t)=\int^{\infty}_{0}d\omega\frac{J(\omega)}{\omega^{2}}\left [\coth(\beta\hbar\omega/2)\cos(\omega t)-i\sin(\omega t)\right ].
\end{align}
The relevant phonon spectral function $J(\omega)$ describes LA (longitudinal acoustic) phonon
coupling via a deformation potential~\cite{krum}, and  is   given by $J(\omega)=\alpha_{p}\omega^{3}\exp(-\omega^{2}/2\omega_{b}^{2})$ where $\alpha_{p}$ is the exciton-phonon coupling strength and $\omega_{b}$ is the phonon cutoff frequency. 


Transforming back to the Schr\"odinger picture and changing variables from $t-t^{\prime}=\tau$, we note that $G_{m}(-\tau)=G_{m}^{*}(\tau)$, and simplify Eq.~(\ref{sec3eq5}) as follows,
\begin{align}
\label{sec3eq12}
&\left . \frac{\partial \rho (t)}{\partial t} \right |_{\rm TC} =\frac{1}{i\hbar}\left [H_{S }^{\prime},\rho(t)\right ] \nonumber \\
&\ \ \ -\frac{1}{\hbar^{2}}\int^{t}_{0}d\tau\sum_{m=g,u}G_{m}(\tau){X}_{m}e^{-iH_{S}^{\prime}\tau/\hbar}{X}_ {m} e^{iH_{S}^{\prime}\tau/\hbar}\rho(t) \nonumber \\
&\ \ \ +\frac{1}{\hbar^{2}}\int^{t}_{0}d\tau\sum_{m=g,u}G_{m}^{*}(\tau){X}_{m}\rho(t)e^{-iH_{S}^{\prime}\tau/\hbar} {X}_{m}e^{iH_{S}^{\prime}\tau/\hbar} \nonumber \\
&\ \ \ +\frac{1}{\hbar^{2}}\int^{t}_{0}d\tau\sum_{m=g,u}G_{m}(\tau)e^{-iH_{S}^{\prime}\tau/\hbar}{X}_{m}
e^{iH_{S}^{\prime}\tau/\hbar}\rho(t){X}_{m} \nonumber \\
&\ \ \ -\frac{1}{\hbar^{2}}\int^{t}_{0}d\tau\sum_{m=g,u}G_{m}^{*}(\tau)\rho(t)
e^{-iH_{S}^{\prime}\tau/\hbar}{X}_{m}
e^{iH_{S}^{\prime}\tau/\hbar}{X}_{m}.
\end{align}
This form of the TC ME   is local in time and can be rewritten in  compact form,
\begin{align}
\label{eq:ME1a}
\left . \frac{\partial \rho(t)}{\partial t} \right |_{\rm TC} &=\frac{1}{i\hbar}\left [H_{S}^{\prime},\rho(t)\right ]+{\cal L}(\rho)
-\frac{1}{\hbar^{2}}\int^{t}_{0}d\tau\sum_{m=g,u}  \nonumber \\
& \left ( G_{m}(\tau) [{X}_{m},e^{-iH_{S}^{\prime}\tau/\hbar}{X}_{m}
e^{iH_{S}^{\prime}\tau/\hbar}\rho(t)]
+ {\rm H.c.} \right ),
\end{align}
where we have added the  ${\cal L}(\rho)$ superoperators  to take into account 
additional dissipation processes that are necessary to obtain a complete description of the system-bath dynamics~\cite{zubairy}. 
We model these additional dissipation processes through the standard Lindblad operators:
\begin{eqnarray}
\label{sec2eq34}
{\cal L}(\rho)&=&\frac{\tilde{\gamma}_{x}}{2}(2{\sigma}^{-}\rho{\sigma}^{+}
-{\sigma}^{+}{\sigma}^{-}\rho-\rho{\sigma}^{+}{\sigma}^{-}) \nonumber \\
&+&\kappa(2{a}\rho{a}^{\dagger}-{a}^{\dagger}{a}\rho-\rho{a}^{\dagger}{a}) \nonumber \\
&+&
\frac{{\gamma}'}{2}(2{\sigma}_{11}\rho{\sigma}_{11}
-{\sigma}_{11}{\sigma}_{11}\rho-\rho{\sigma}_{11}{\sigma}_{11}),
\end{eqnarray}
where $\sigma_{11}= \sigma^+  \sigma^-$, $2\kappa$ is the cavity decay rate, $\gamma'$ is the pure dephasing rate, and  $\tilde{\gamma}_{x}=\gamma_{x}\langle B\rangle^{2}$ is the radiative decay rate~\cite{roy_hughes_prx,roy}. 
Note that we 
include additional pure dephasing effects beyond the IBM, e.g., perhaps caused by  quadratic QD-acoustic phonon and phonon-phonon interactions~\cite{quadratic_phonon,Rudin:PRL06}, with a rate $\gamma^{\prime}$;  this pure dephasing contribution to the  ZPL broadening is also known to increase linearly with increasing temperatures~\cite{besombes,BorriPRL:2001}.

The timescales associated with the phonon and radiative processes are substantially different~\cite{roy}. Phonon processes happen at picosecond timescales whereas radiative timescales are smaller  by (typically) at least an order of magnitude~\cite{kaer}. This results in a substantial simplification of the TC ME derived above. Extending the upper limit of the integral ($t\rightarrow\infty$), Eq.~(\ref{eq:ME1a}) can be rewritten as
\begin{align}
\label{eq:ME1b}
\left . \frac{\partial \rho(t)}{\partial t} \right |_{\rm TC}& =  \frac{1}{i\hbar}\left [H_{S}^{\prime},\rho(t)\right ]+{\cal L}(\rho)-\frac{1}{\hbar^{2}}\int^{\infty}_{0}d\tau\sum_{m=g,u}
 \nonumber \\
& \left (G_{m}(\tau)[{X}_{m},e^{-iH_{S}^{\prime}\tau/\hbar}{X}_{m}
e^{iH_{S}^{\prime}\tau/\hbar}\rho(t)]
+{\rm H.c.} \right). 
\end{align}
This ME equation  is now in a TC Markov form. The validity of extending the time limits of the integration to $\infty$ depends  on the ensuing system dynamics introduced by the external drive, and the exciton-cavity coupling (through $g$), with respect to the phonon bath
relaxation time. For instance, a cavity-QED system driven by a short pulse may require a fully non-Markovian treatment to obtain a correct description of nonradiative processes. Non-Markovian effﬀects within the polaron frame ME are likely to affect the population dynamics only at very short times and the difference between non-Markovian and Markovian dynamics vanish at longer timescales~\cite{nazir2}. For the present system of interest (with cw drives), the Markov approximation is found to be an excellent approximation as will be shown in the next section. 

Using similar steps as above, Eq.~(\ref{sec_new_eq1}) leads to a ME of the NZ form that involves integrating the system dynamics over all previous times~\cite{imamoglu,carmichael}, so that
\begin{align}
\label{eq:ME2}
&\left . \frac{\partial \rho(t)}{\partial t}\right |_{\rm NZ}=\frac{1}{i\hbar}\left [H_{S}^{\prime},\rho(t)\right ]+{\cal L}(\rho)
-\frac{1}{\hbar^{2}}\int^{t}_{0} d\tau\sum_{m=g,u}  \nonumber \\
&\left (G_{m}(\tau)[ X_{m}, e^{-iH_{S}^{\prime}\tau/\hbar} X_{m}\rho(t-\tau)
e^{iH_{S}^{\prime}\tau/\hbar}]+{\rm H.c.}
\right ). 
\end{align}
This ME is non-local in time and generates equations of motion which are convoluted in time.

We note that the validity of the Born approximation, used to derive the above MEs,  depends on the physical parameters of the cavity-QED system. A general condition for the validity of the second-order Born approximation
has been derived by McCutcheon and Nazir~\cite{nazir2} and is given by $\frac{\eta_{x}^{2}}{\omega_{b}^{2}}(1-\langle B\rangle^4)\ll 1$. 
A similar condition for the validity of the Born approximation for the cavity-QED system can easily be derived with relevant numbers well below unity (since $\eta_x\ll \omega_b$ and $g \ll \omega_b$, with $\omega_b=1~$meV for this work); 
this suggests that a second-order Born approximation is perfectly valid for our present system. The main reason for this observation is that  nonperturbative coupling effects beyond second-order Born are already captured within the polaron transform itself.

One can also derive a ME describing the cavity-QED dynamics  without using the polaron transformation~\cite{nazir1}. One common approach  is to use a standard {\em weak-coupling theory} which only includes exciton-phonon interactions to second-order and is known to breakdown for higher temperatures and strong exciton-phonon couplings~\cite{nazir2}; with cavity coupling, these weak phonon-coupling failures can become even stronger~\cite{roy_hughes,roy_hughes_prx}. A more general approach using a variationally-optimized ME has also been described  
by  McCutcheon {\em et al.}~\cite{nazir_new}. The variational ME includes nonperturbative phonon effects and multiphonon processes over a wide range of parameters (especially important for very large driving fields).
Gl\"assl {\em et al.}~\cite{axt_new} have  gone beyond a polaron picture as well, and carried out  numerically-exact real-time path integrals and a fourth-order correlation expansion
to model the long-time dynamics and the stationary nonequilibrium state of an optically driven QD (coupled to acoustic phonons).
For the purposes of our paper, as noted above, the polaron transformation is rigorously valid  as the coherent laser pump rates and the exciton-cavity coupling rate are much smaller than the phonon cutoff frequency~\cite{nazir_new}. Moreover, a clear 
advantage of our approach is the ease with which cavity coupling is included,
and, for example, the well known Jaynes-Cummings model is formally recovered with phonon coupling switched off~\cite{imamoglu}.


With coherent pumping, we  define the incoherent fluorescence spectra associated with cavity mode emission ($S_{c}$), and QD emission ($S_{x}$) \cite{roy}, as follows:
\begin{align}
S_{c}(\omega) & \propto\frac{\kappa}{\pi}\lim_{t\rightarrow\infty}
{\rm Re}\left[\int^{\infty}_{0}d\tau\, (\langle{a}^{\dagger}(t+\tau){a}(t)\rangle \nonumber \right . \\
&-\left . \langle{a}(t)\rangle\langle{a}^{\dagger}(t)\rangle) e^{i(\omega_{L}-\omega) \tau} \right ], \\ \nonumber
S_{x}(\omega) & \propto\frac{\tilde{\gamma}_{x}}{\pi}\lim_{t\rightarrow\infty}{\rm Re}\left [\int^{\infty}_{0}d\tau\, (\langle{\sigma}^+(t+\tau){\sigma}^-(t)\rangle \right .\\
&-\left . \langle{\sigma}^{+}(t)\rangle\langle{\sigma}^{-}(t)\rangle) e^{\phi^{*}(\tau)}e^{i(\omega_{L}-\omega) \tau} \right ].
\label{spec:defa}
\end{align}
We highlight the phase term $e^{\phi^{*}(\tau)}$ appearing in the definition of the exciton spectra, $S_{x}(\omega)$, which arises due to the polaron transformation and accounts for the dephasing of the exciton coherence in the phonon reservoir. In the polaron frame, the exciton  autocorrelation function $\langle{\sigma}^{+}(t+\tau){\sigma}^{-}(t)\rangle$ is transformed to $\langle{\sigma}^{+}(t+\tau){B}_{+}(t+\tau){\sigma}^{-}(t){B}_{-}(t)\rangle$. The trace over the phonon degrees of freedom and the QD Hilbert space is  easily evaluated as the Hilbert space is separable in the polaron frame. 

It is also useful to define the QD susceptibility (or polarizability), $\chi(\omega)$, that describes the linear response of the QD in the presence of phonon coupling and phenomenological dissipation only (i.e., not coupled to the cavity mode, $g\simeq 0$, and driven by a linear optical field). This {\em bare} polarizability function is obtained from
\begin{align}
\chi(\omega)&=\chi^{\prime}(\omega) + i \chi^{\prime \prime}(\omega)\nonumber \\
\chi(\omega)&\propto\int^{\infty}_{0}d\tau\langle{\sigma}^{-}(\tau){\sigma}^{+}(0)\rangle e^{\phi^{*}(\tau)}e^{i(\omega_{x}-\omega) \tau},
\label{spec:defb}
\end{align}
where we now use a rotating frame with respect to $\omega_x$.
This simple looking form for $\chi(\omega)$, with the phonon contribution described by the phonon correlation function $\phi(t)$, is due to the polaron transformation. To obtain the above QD response function, the expectation is evaluated in the absence of any exciton-cavity coupling,
and we excite the QD exciton with a linear (weak)
incoherent pump process, via the Lindblad superoperator:
${P_{x}}(2{\sigma}^{+}\rho{\sigma}^{-}-{\sigma}^{-}
{\sigma}^{+}\rho-\rho{\sigma}^{-}{\sigma}^{+})$ with a suitably small $P_x$.
%
The linear absorption spectra is   defined through $\chi^{\prime\prime}(\omega)$.  

It is interesting to recognize that the phonon correlation function $\phi(t)$ is explicitly present in the definition of the $S_{x}$ and $\chi(\omega)$. The dynamics of equations we solve
is effectively Markovian in the polaron frame, which is much easier to work with from a numerical viewpoint. However,  the transformation back to the original frame reintroduces the non-Markovian phonon dynamics through the phonon correlation function $\phi(t)$. Note also that the cavity spectrum does not depend directly on $\phi(t)$, so the cavity spectrum may not depend at all on any non-Makovian dynamics (at least from a mathematical viewpoint, and if one works in a polaron frame).


The fluorescence spectra are calculated by the numerical solution of the  ME driven by a steady-state pump $\eta_x$. The ME is solved in a Jaynes-Cummings basis with states $\ket{0}$, $\ket{1L}$, $\ket{1U}$, $\ket{2L}$, $\ket{2U}$ and so on. Denoting the photon states $\ket{n}$, with $n=0,1,2,...,$ and exciton states $\ket{e/g}$, the Jaynes-Cummings basis states are given by $\ket{0}=\ket{g}\ket{0}$,$\ket{1L}=1/\sqrt{2}(\ket{g}\ket{1}-\ket{e}\ket{0})$,
$\ket{1U}=1/\sqrt{2}(\ket{g}\ket{1}+\ket{e}\ket{0})$, $etc$. The system is initialized with the exciton and cavity in the ground state which then evolve, in the presence of a cw pump, to a dynamical steady state. Using the quantum regression formula~\cite{me_4}, we subsequently compute the relevant two-time correlation function whose decay is initialized by the steady state population and coherences. The  calculations are performed, in part,  using the {\em quantum optics toolbox} by Tan~\cite{QOToolbox}.  We note that, in general, it is vital to include multiphoton effects to get the correct (numerically converged in the photon number)  fluorescence spectra; though a one photon basis is (obviously) enough to obtain the linear absorption. We have verified the need for multiphoton effects
elsewhere  and find that the truncation of the hierarchy of the photon states to two-photon-correlations (two photon states) is sufficient (and necessary)~for exciton driving \cite{roy_hughes}. A broader discussion of the role of multiphoton processes, especially in the context of intensity power broadening, is presented in Ref.~\onlinecite{roy_hughes_prx}.

\begin{figure}[t]
\centering\includegraphics[trim = 5mm 0mm 0mm 0mm, clip, width=.99\columnwidth]{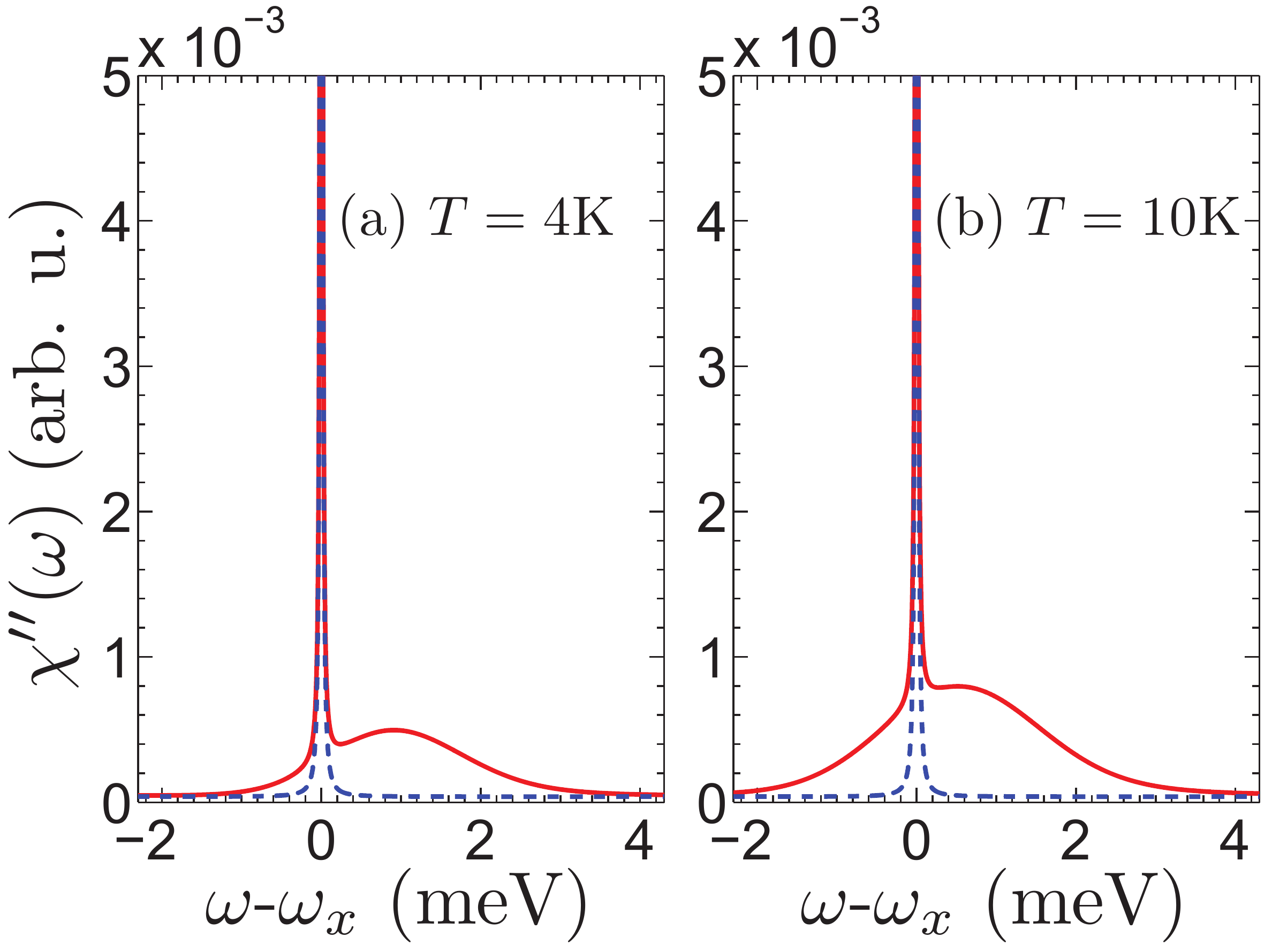}
\caption{ (Color online)  Linear absorption spectra of the QD, $\chi^{\prime\prime}(\omega)$, for two different phonon bath temperatures: (a) $T=4$~K  and (b) $T=10$~K. The  red (solid) line includes phonon coupling and blue (dashed) line has no phonon coupling. The QD decay parameters  $\tilde{\gamma}_{x}=1~\mu$eV and $\gamma^{\prime}=2~\mu$eV are used in both plots, which give the broadening of the ZPL. The asymmetry in the absorption spectra due to the phonon reservoir is clearly seen, especially at lower temperatures.} 
\label{fig:linearspec}
\end{figure}

\section{Numerical Results}
\label{results}
In this section, we present and discuss our numerical results.
For all calculations that follow, we use the following phonon parameters: 
$\omega_{b}=1~$meV and $\alpha_{p}/(2\pi)^2=0.06~{\rm ps}^{2}$, which are typical numbers for InAs/GaAs QDs~\cite{JiaoJPC2008,hughes1,roy_hughes}.
For the cavity decay rate, we choose $\kappa=50~\mu$eV, and for the 
exciton decays, $\tilde{\gamma}_{x}=1~\mu$eV and $\gamma^{\prime}=2~\mu$eV. For simplicity, we will fix the pure dephasing rate below, but this rate is also known to increase linearly with temperature~\cite{BorriPRL:2001,ota}. The other parameters are generally varied or given in the figure captions. 


\subsection{Linear absorption spectra}
In Fig.~\ref{fig:linearspec} we show the linear absorption spectra of the QD exciton, $\chi^{\prime\prime}(\omega)$, for two different temperatures of the phonon bath. The absorption spectra in the vicinity of the QD resonance is dominated by the IBM spectral function which allows the dot to absorb photons by phonon-assisted processes. In the absence of any phonon coupling (blue dashed lines in Fig.~\ref{fig:linearspec}), the absorption spectrum is a simple Lorentzian whose width is determined by radiative broadening and pure dephasing (the ZPL linewidth). The asymmetry in the absorption spectra in the presence of phonons is clearly seen at low temperatures, which is caused by different probabilities to absorb or emit a  phonon; i.e., the latter process is more likely at lower temperatures. 


\begin{figure}[t]
{\includegraphics[width=.99\columnwidth]{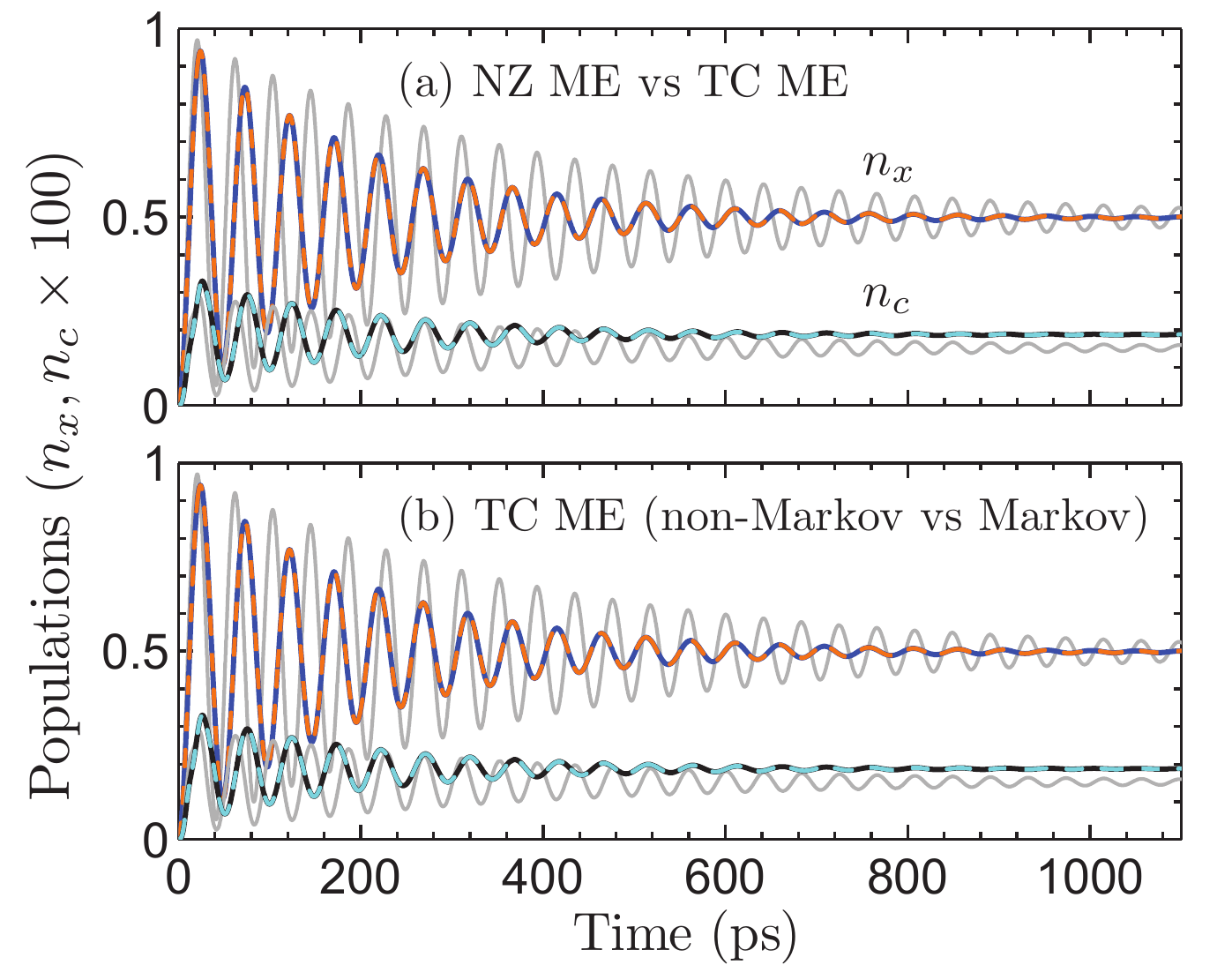}}\\
\caption{ (Color online) (a) Comparison of the population dynamics generated by the NZ and the TC form of the ME, both to second-order in the system-phonon--resorvoir coupling. The upper blue (dark solid) line is the QD  exciton population $n_{x}(t)$ using the NZ solution, and the orange (light dashed) line is the TC ME solution. We also plot the cavity mode population $n_{c}(t)$. The grey (light solid) lines show the  exciton and the cavity mode populations in the absence of any microscopic phonon coupling (i.e., only including ZPL broadening and cavity coupling). In (b) we compare the population dynamics of the TC ME and the TC Markov    ME. The blue (solid) line shows the exciton population $n_{x}(t)$ using the TC ME, while the orange (dashed) line is the dynamics generated by the TC Markov ME form.  All the plots were obtained by truncating the hierarchy of equations of motion at the one-photon limit, and using  $T=10~$K, 
$\tilde{\gamma}_{x}=1~\mu$eV, $\kappa=50~\mu$eV, $\gamma^{\prime}=2~\mu$eV, $g=20~\mu$eV,  and $\eta_x=0.05~$meV. 
\\  
}
\label{fig:master1}
\end{figure}
\begin{figure}[t!]
{\includegraphics[width=.99\columnwidth]{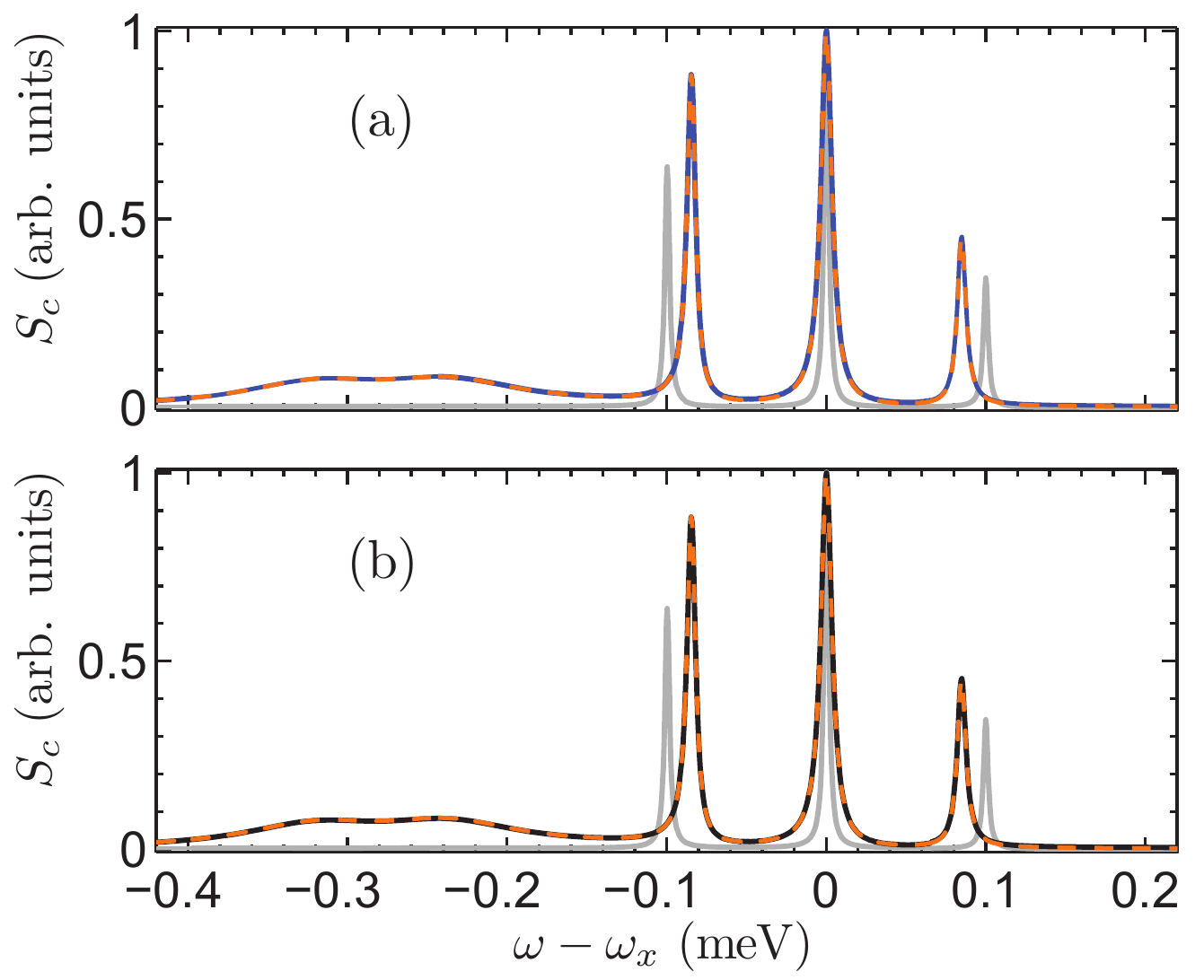}}
\caption{
(Color online) (a) Corresponding cavity-emitted resonance fluorescence spectra from Fig.~3.  The blue (solid) line is the cavity-emitted fluorescence spectrum ($S_c$) computed using the NZ form and the orange (dashed) line is the spectrum generated by the TC ME. The grey solid lines show the spectrum in the absence of any phonon coupling (apart from the ZPL). In (b) we compare the cavity-emitted  spectra generated by the TC and TC Markov ME results. 
}
\label{fig:master2}
\end{figure}

\subsection{TC vs NZ master equation solutions: population dynamics and Mollow triplet spectra}
In Fig.~\ref{fig:master1} we compare the population dynamics generated by the NZ ME [Eq.~(\ref{eq:ME2})],  the TC ME [Eq.~(\ref{eq:ME1a})] and the TC Markov ME [Eq.~(\ref{eq:ME1b})].
We excite resonantly (i.e., $\omega_L=\omega_x$), with the cavity detuned to lower
energies ($\omega_x-\omega_c=0.28~$meV), using
the exciton pump rate, $\eta_x=0.05~$meV.
As can be seen from the plots,  all three ME forms produce identical results for the evolution of the cavity  and QD populations, defined as $n_{c}(t)=\langle{a}^{\dagger}{a}(t)\rangle$ and $n_{x}(t)=\langle{\sigma}^{+}{\sigma}^{-}(t)\rangle$. We also compare the cavity-emitted fluorescence spectra ($S_c$) as shown in Fig.~\ref{fig:master2} (defined in the previous section), generated by the respective MEs, which are again found to be identical (or, at least, indistinguishable). Hence, in the remainder of the paper we use the TC Markov  form of the ME as it is much easier to work with and is justified here to be an excellent approximation.
Note that for the bath temperature of 10~K, $\braket{B} = 0.85$, which reduces
the effect of both $\eta_c$ and $g$, and causes a smaller  triplet  splitting. 

For numerical convenience, the calculations\ in this subsection (shown in  Fig.~\ref{fig:master1} and Fig.~\ref{fig:master2}) were obtained by truncating the hierarchy of equations of motion at the one photon limit because of the computational complexity of solving the  NZ ME form. However, all the results discussed in the subsequent sections use the TC Markov ME with the required (i.e., numerically converged) number of multiphoton processes; so the fluorescence spectrum will be seen to qualitatively change, especially near the cavity resonance (see Ref.~\onlinecite{roy_hughes} for more details).

\begin{figure}[b!]
\centering\includegraphics[width=.99\columnwidth]{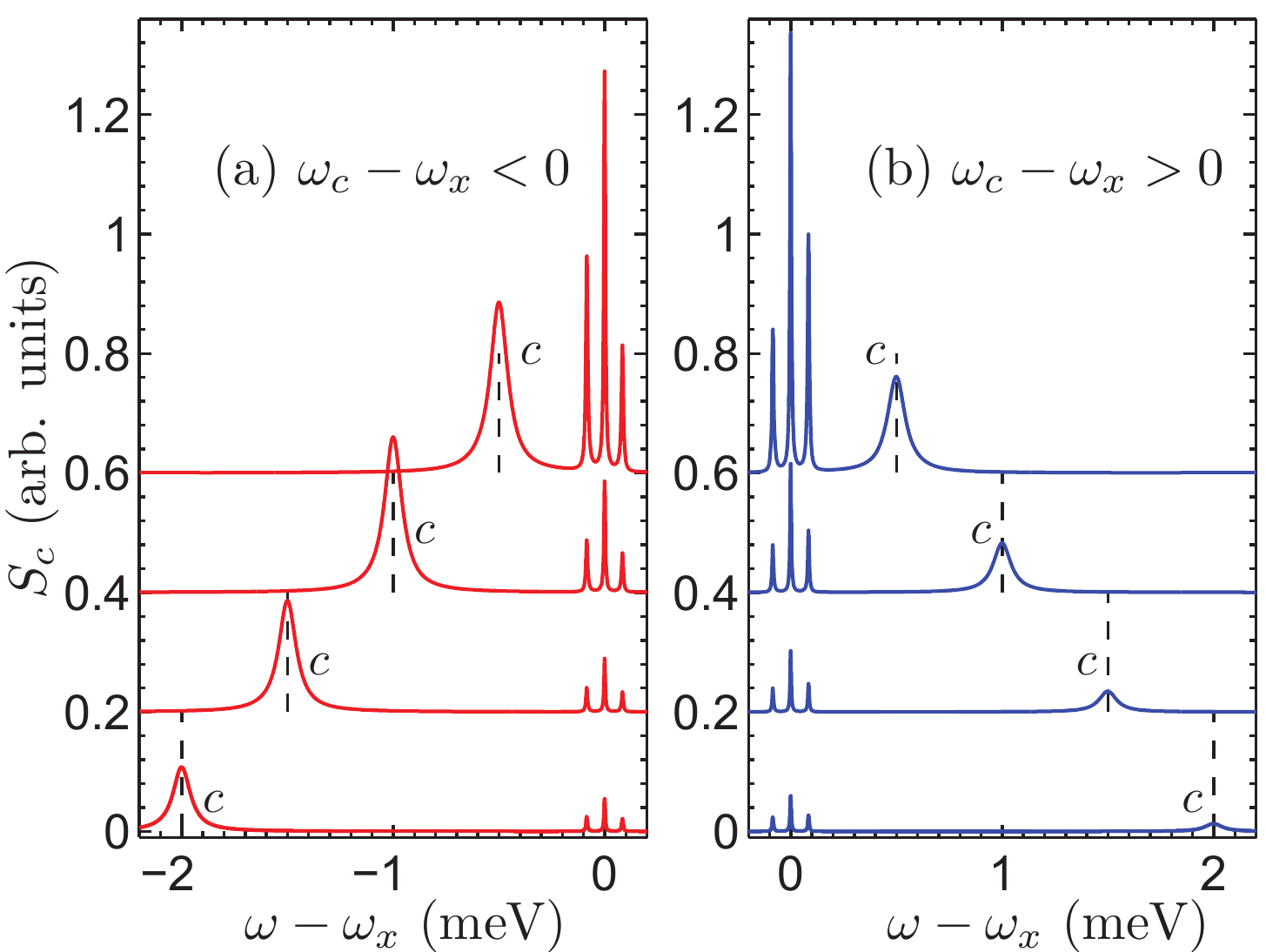}
\caption{(Color online) Detuning dependence of $S_c$  for different exciton-cavity detunings.  The other parameters are the same as in Fig.~\ref{fig:master1}, again with the phonon bath temperature at $T=10~$K. In (a) the red lines denote $S_c$ when the cavity is detuned to the left of the QD (exciton). In (b) the blue lines denote the same spectra when the cavity is detuned to the right of the QD. Note that all the spectra are normalized relative to the first spectrum on the top left.}
\label{fig:detuning1}
\end{figure}

\subsection{QD-cavity detuning dependence of the phonon-dressed Mollow triplet}
In Fig.~\ref{fig:detuning1} we study the detuning dependence of $S_c$  for different QD-cavity detunings. The fluorescence spectra are clearly asymmetric with respect to the cavity mode detuning; this behavior is drastically different to simple Lorentzian-decay models which would yield symmetric Mollow spectra, i.e., with respect to the exciton-cavity detuning. Cavity  emission is  considerably enhanced when the QD is detuned to the right of the cavity mode as the cavity-QD coupling is further strengthened by phonon emission processes. At a detuning of $\pm\,$1~meV, the red-detuned cavity peak is found to be about double that of the blue detuned cavity
mode peak, which is consistent with experimental data using incoherent excitation, e.g., as shown
by Hennessy {\em et al.}~\cite{HennessyNature:2007}.
  Also note that the cavity-mode peak  initially increases, relative to the Mollow triplet, as the QD exciton is red detuned from the exciton;  but eventually decreases  for sufficiently large QD-cavity detunings. This suggests that the coupling between the QD and the cavity is optimum at a particular detuning and is determined by the QD-phonon coupling strength and the intrinsic spectral properties of the phonon reservoir. 
This observation is significant for experimentalists as it suggests that off-resonance cavity-QED  physics is best studied at a particular QD-cavity detuning where the mode emission is maximized. Note that the emission at the cavity mode frequency is eventually suppressed at very large QD detunings $\Delta_{cx}\simeq\pm5$ meV, which is in agreement with recent experimental observations on site-controlled QDs in cavities~\cite{hughes2}; however, these measurements were again carried
out using incoherent excitation, so not in the domain of resonance fluorescence.  The maximum coupling between the QD and the cavity also depends on temperature, e.g., the QD-cavity detuning at which the QD-cavity mode coupling is maximum increases with temperature.

It is informative to note that the detuning dependence of the spectra closely follows the spectral characteristics of the phonon correlation function. An intuitive way to understand this detuning dependence is using effective phonon rates in a Lindblad-type description~\cite{roy_hughes_prx}, which are similar to the cavity-feeding rates introduced in Refs.~\onlinecite{HohenesterPRB:2009,HohenesterPRB:2010}. These phonon scattering rates are related to the Fourier component of the phonon correlation function at the QD-cavity mode detuning. The same type of feature can be seen in the asymmetry of the sidebands of the vacuum-Rabi doublet in an incoherently pumped strongly coupled QD-cavity system~\cite{hughes2}; the asymmetry of the doublet depends closely on the spectral characteristics of the phonon correlation function. Hence, these features in the fluorescence spectra and linear absorption reveal subtle details of the underlying phonon reservoir, which can be probed as a function of the exciton-cavity detuning. The advantage of coherently excited systems is that they are much cleaner to study, since for incoherent excitation one never really knows the details of the 
pumping scenario (which may involve a number of exciton states and contributions from different QDs).

\begin{figure}[b]
\centering\includegraphics[width=0.99\columnwidth]{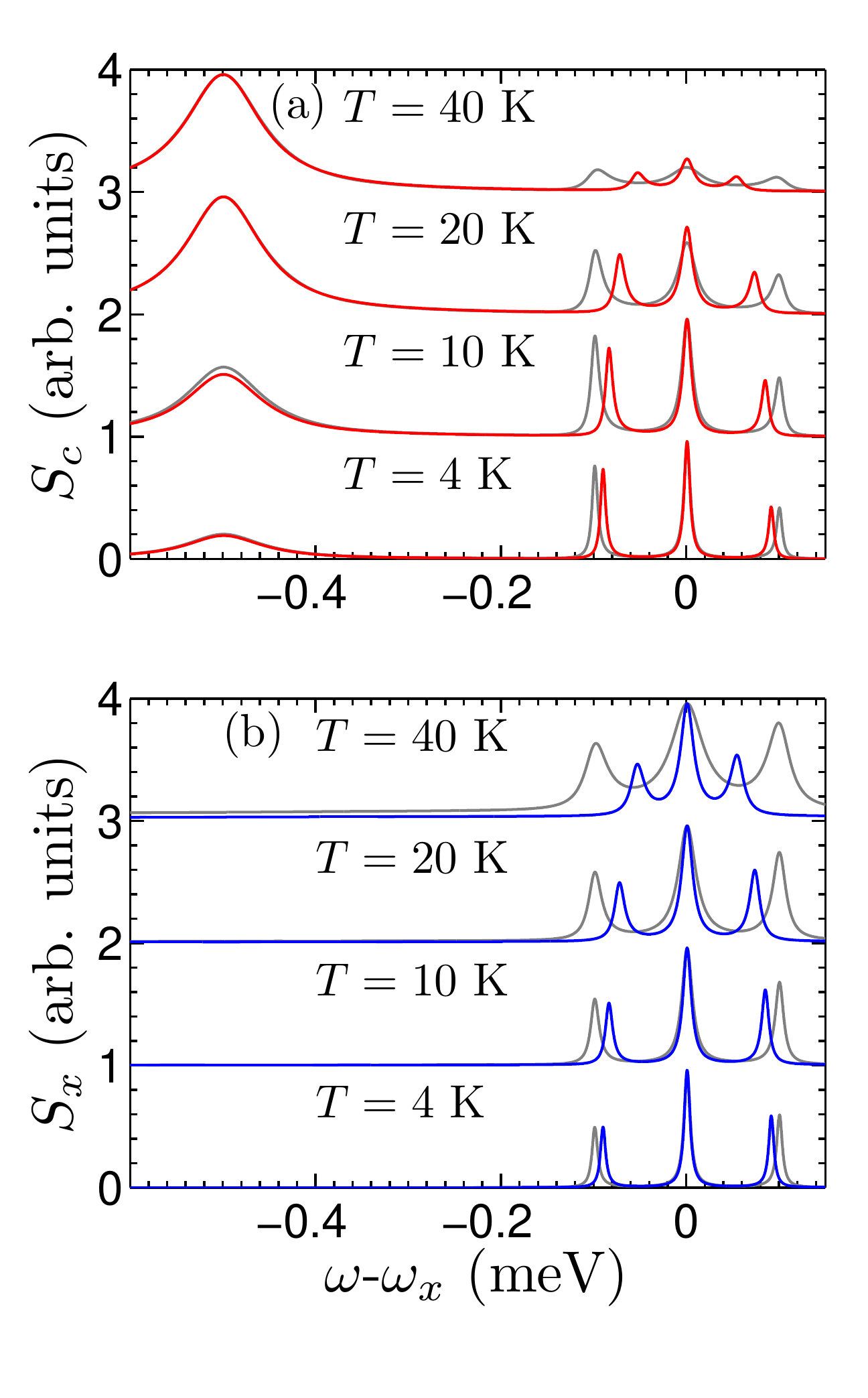}
\caption{(Color online)  Temperature dependence of the (a) cavity-emitted resonance fluorescence spectra (red, dark line) and (b) exciton-emitted resonance fluorescence spectra (blue, dark line) for different temperatures of the phonon reservoir. 
Also shown is the  one-phonon calculations (grey, light curves). The QD-cavity detuning, $\Delta_{cx}=-0.5$~meV, with the phonon reservoir temperatures: $T=4$~K, $T=10$~K, $T=20$~K, and $T=40$~K.  The other parameters are the same as in Fig.~\ref{fig:master1}.
}
\label{fig:temp}
\end{figure}

%
%

\begin{SCfigure*}
\includegraphics[scale=0.45]{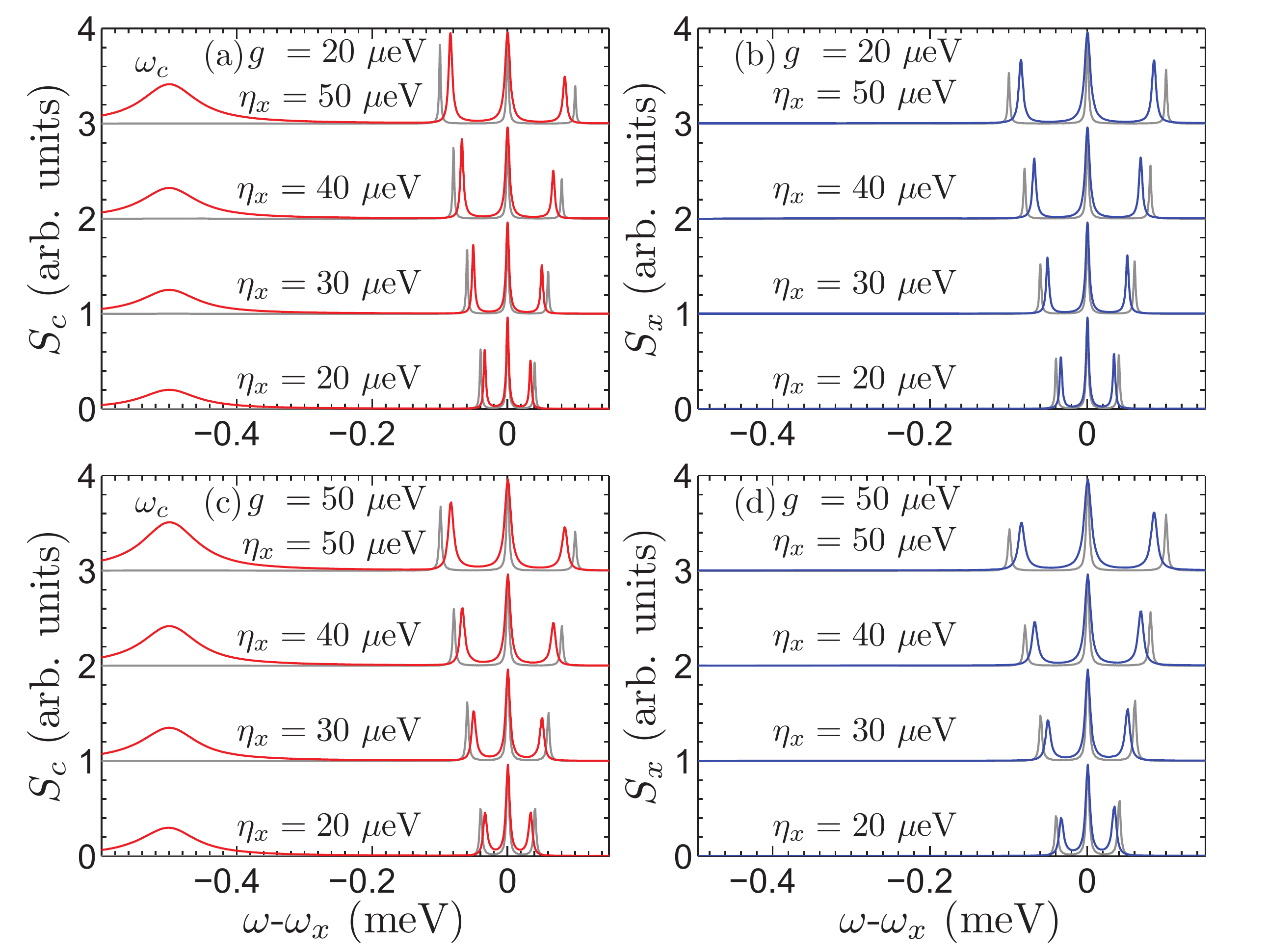}
 \caption{(Color online) 
  Resonance fluorescence spectra for various cw-laser drives and two different exciton-cavity coupling $g.$ The phonon reservoir at $T=10$~K, and we use $\Delta_{cx}=-0.5$~meV, $\tilde{\gamma}_{x}=1~\mu$eV, $\kappa=50~\mu$eV, and $\gamma^{\prime}=2~\mu$eV. In Figs.~(a) and (b) we plot 
$S_c$ and $S_x$ for exciton-cavity coupling $g=20~\mu$eV with various cw-laser driving $\eta_{x}$ which are shown as red (dark) lines and blue (dark) lines, respectively (left and right panels). Also plotted are the grey (solid) lines which are the corresponding fluorescence spectra in the absence of any phonon coupling (apart from the inclusion of the ZPL broadening). In Figs.~(c) and (d) we plot $S_c$  and $S_x$  with the larger  $g=50~\mu$eV using similar parameters as in Figs.~(a) and (b).}
  \label{fig:spectra}
\end{SCfigure*}

\subsection{Temperature dependence of the resonance fluorescence spectra}
In Fig.~\ref{fig:temp} we investigate the temperature dependence of the resonance fluorescence spectra and also analyse the role of one-phonon scattering versus multiphonon scattering. In Fig.~\ref{fig:temp}(a), we show the red-detuned cavity mode, where the grey (solid) lines denote the one-phonon limit of the full polaron cavity-emitted spectra which is shown with the red (dark) line. We observe that the emission at the cavity mode relative to the emission at the QD frequency increases with temperature. 
In Fig.~\ref{fig:temp}(b), we plot the QD exciton spectra, where
the full-polaron QD 
spectra is now plotted as the blue (dark) lines. 

The one-phonon limit of the full-polaron ME is obtained by a perturbative expansion of the relevant phonon correlation functions to lowest order; namely, we approximate $G_{g}(\tau)\simeq 0$ and $G_{u}(\tau)\simeq \phi(t)$. The TC Markov ME  is then solved numerically for the above choice of phonon correlation functions to study the fluorescence spectra in the one-phonon limit. In the weak phonon coupling theory, $\langle B\rangle\simeq 1$, suggesting that one-phonon processes do not renormalize the Rabi frequency of the cw-laser drive, $\eta_{x}$, or the QD-cavity coupling constant $g$. 
The discrepancy between the one-phonon and full-polaron solution of the TC  ME increases with temperature, and the one-phonon solution is found to {\em overestimate} the EID in comparison with the full polaron solution. 
These results are consistent with the findings of McCutcheon and Nazir~\cite{nazir2}, who adopt a polaron ME for analysing QD Rabi oscillations in the presence of acoustic phonon scattering (with no cavity).
Note that since the one-phonon truncation for the full polaron theory does not renormalize the  laser Rabi frequency, the locations of the triplets are identical to that for the case of no phonons. Hence a full polaron theory can be essential to find the correct center frequencies of the Mollow triplet resonances. Moreover, the discrepancy between the sideband locations increases with temperature. However, the weak phonon theory
can obtain accurate simulations of the cavity resonance, depending upon the exciton-cavity detuning.
Without any acoustic phonon coupling, the Mollow spectra
are significantly different, e.g., there is no EID
and there is negligible oscillator strength at the cavity mode.
These results suggest that a simple atomiclike ME  would drastically fail in the present semiconductor cavity-QED excitation regime.

\begin{SCfigure*}
\includegraphics[scale=0.45]{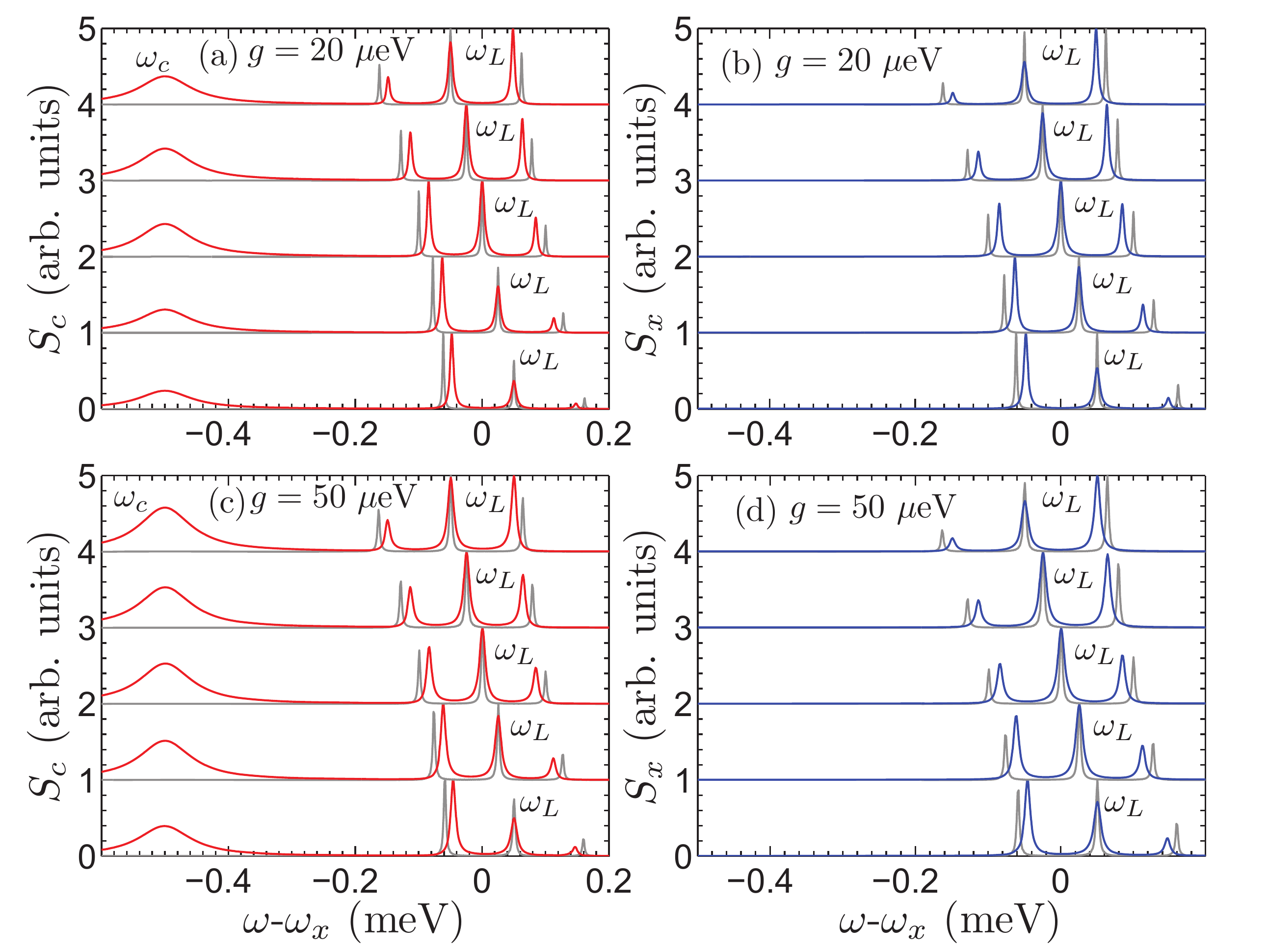}
\caption{(Color online) Similar to  Fig.~\ref{fig:spectra}, but 
with $\eta_x$ fixed at $50~\mu$eV and for different exciton-cw laser detunings $\Delta_{Lx}$. In Figs.~(a) and (b) we plot the resonance fluorescence spectra ($S_c$ and $S_x$, respectively) for QD-cavity coupling $g=20~\mu$eV with various exciton-laser detunings respectively plotted as red (solid) lines and blue (solid) lines.  Also shown are the fluorescence spectra in the absence of any phonon coupling as grey-solid lines (apart from the ZPL). In Figs.~(c) and (d) we  plot the fluorescence spectra for exciton-cavity coupling $g=50~\mu$eV with identical detunings and cw laser drives as in Figs.~(a) and (b).}
\label{fig:detuning}
\end{SCfigure*}

\subsection{Resonance fluorescence spectra versus  exciton pump rate}
In Fig.~\ref{fig:spectra} we display the normalized resonance fluorescence spectra for various cw laser drives and dot-cavity couplings, using
a fixed QD-cavity detuning. The temperature of the bath is  at  $T=10~$K. From Fig.~\ref{fig:spectra}(a) and Fig.~\ref{fig:spectra}(c), we observe that the relative emission at the cavity mode frequency increases with increasing drive which  is caused by the enhanced coupling between the cavity mode and the QD, as the low energy triplet component is pushed closer to the cavity mode frequency. The Mollow triplet peak which is proximal to the cavity mode has a higher spectral weight than the distant component which is due to stronger relative coupling with the cavity mode.  However, we note that the asymmetry in the spectral weights decreases for larger $g$ as both the proximal and the distant component of the triplet are more or less equally coupled to the cavity mode.
For increasing pump strengths, 
it is clear that the linewidth broadening is larger in the presence of phonons.
 Further increased broadening of the Mollow sideband occurs with increasing $g$~\cite{roy_hughes}.

Apart from the lack of any emission at the mode frequency in the QD emission spectra, the asymmetry in the Mollow triplets in $S_x$  is substantially smaller than $S_c$. Furthermore, the high energy component of the triplet has a larger relative spectral weight than the low energy component. This asymmetry in the sidebands spectral weights increase with increasing $g$ and cw-laser drive $\eta_{x}$. Also, the asymmetry of the sidebands are not mainly due to phonon-induced effects as they are present even in the absence of phonons; thus these features are primarily cavity coupling effects and are easy to understand in terms of relative coupling with the cavity mode. The low-energy component of the triplet is closer to the cavity mode and couples more strongly with it.

\subsection{Fluorescence spectra with an off-resonance cw drive}
In Fig.~\ref{fig:detuning} we study $S_c$ and $S_x$ for different exciton-laser detunings, using  two different exciton-cavity couplings, $g$. Again we fix the phonon reservoir at $T=10$~K. We first note that, unlike the on-resonance spectra (see Fig.~\ref{fig:spectra}), the relative spectral weights of the Mollow triplets are also now determined by $\Delta_{Lx}$ (the exciton-cw laser  detuning). The high-energy component of the Mollow triplet has larger relative spectral weight if the cw-laser is detuned to the left of the exciton (lower energies). Conversely, the low-energy component of the triplet has larger relative spectral weight if the cw-laser is detuned to the right of the exciton (higher energies). This observation is also consisent with the features of $S_x$. An off-resonant cw-laser drive shifts the entire triplet with a corresponding shift in the location of the sidebands. A positive exciton-cw laser detuning shifts the triplet structure further away from the cavity mode but increases the spectral weight in the low energy sideband as it is pushed closer to the bare exciton resonance. A negative exciton-cw laser detuning shifts the triplet structure towards the cavity mode, and increases the spectral weight in the high energy sideband as it is pushed closer to the bare exciton resonance. Also note the asymmetry in the low-energy sideband of the triplet in the top panel and the high-energy sideband in the bottom panel in both Fig.~\ref{fig:detuning}(a) and Fig.~\ref{fig:detuning}(c) which is explained by the difference in the interaction with the cavity owing to different effective spectral detuning between the respective sideband and the cavity mode. The asymmetry in the spectral weights is suppressed with increasing QD-cavity coupling as seen by comparing, e.g., the relative weights in Fig.~\ref{fig:detuning}(a) and Fig.~\ref{fig:detuning}(c). The enhanced exciton-cavity interaction overrides, to some extent, the effects of the spectral shift induced by the off-resonant drive. Finally, we note that phonon coupling enhances the asymmetry in the relative spectral weights of the Mollow triplets due to enhanced exciton coupling and reduced sideband splitting.

\begin{SCfigure*}
\centering\includegraphics[scale=0.412]{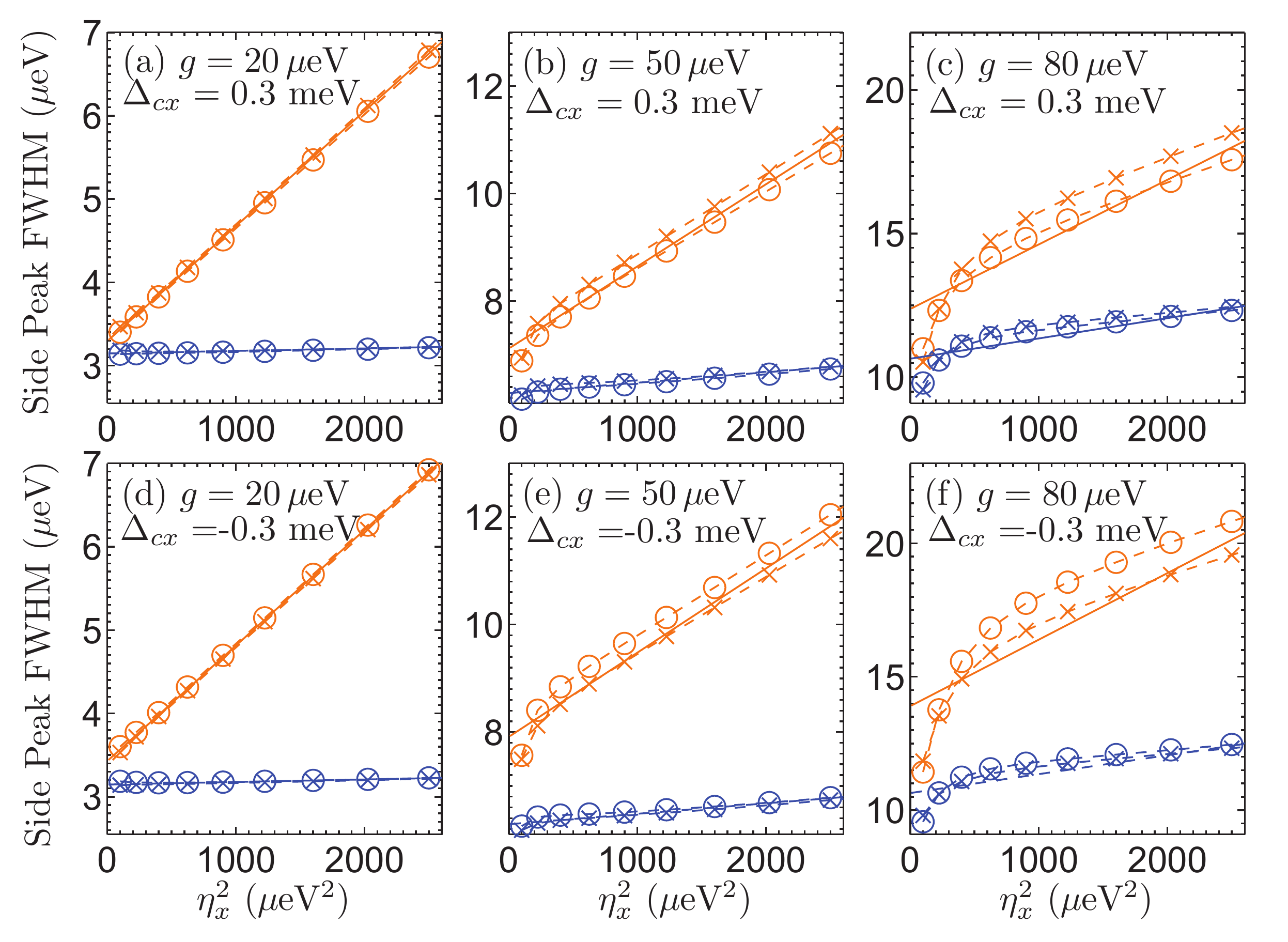}
\caption{(Color online) Numerically extracted FWHM of the on-resonance Mollow triplet spectrum emitted via the cavity mode ($S_c$) for various exciton-cavity detunings and $g$, as a function of the square of the Rabi frequency $\eta_x$. 
In Figs.~(a)-(c) we plot the FWHM of the sidebands of the fluorescence spectrum for $\Delta_{cx}=0.3$~meV. 
The orange (light) data is with  phonon coupling and the blue (dark) data is for no phonon coupling. The crosses denote the high frequency and the circles denote the low frequency component of the Mollow triplet. Also plotted as a solid line with corresponding color is the best linear fit to the data. In Figs.~(d)-(f) we show the FWHM of the sidebands of the fluorescence spectrum for $\Delta_{cx}=-0.3$~meV, using identical parameters as in the above. }
\label{fig:fwhm1}
\end{SCfigure*}

\begin{SCfigure*}
\centering\includegraphics[scale=0.4]{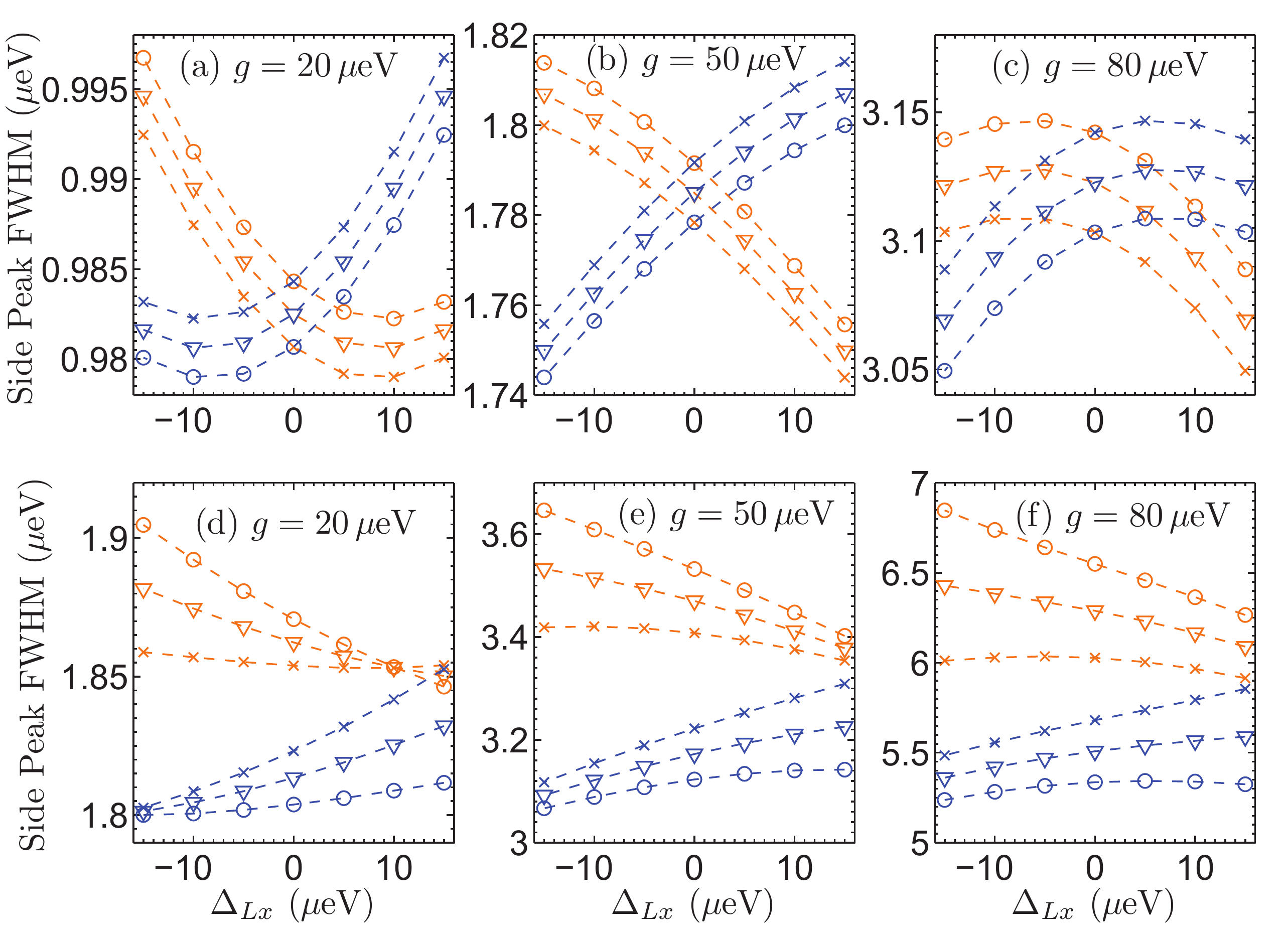}
\caption{(Color online) Numerically extracted FWHM of the off-resonance Mollow triplet spectrum emitted via the cavity mode ($S_c$) for different exciton-cavity detunings and $g$, as a function of the QD-laser detuning $\Delta_{xL}$. 
In Figs.~(a)-(c) we plot the FWHM of the sidebands of the fluorescence spectrum in the absence of phonons with the exciton-cavity couplings: $g=20~\mu$eV, $g=50~\mu$eV and $g=80~\mu$eV. The blue data is with $\Delta_{cx}=0.3$~meV and the red data is for $\Delta_{cx}=-0.3$~meV. The crosses denote the high frequency and the circles denote the low frequency component of the Mollow triplet. The triangles are the average of the two sideband FWHM. In Figs.~(d)-(f) we study the FWHM of the sidebands of the fluorescence spectrum, with phonon scattering at    $T=10$~K. }
\label{fig:fwhm2}
\end{SCfigure*}


\subsection{FWHM (linewidth) of the Mollow sidebands}
We next study the broadening of the Mollow sidebands for various excitation regimes. 

In Fig.~\ref{fig:fwhm1} we show the numerically extracted FWHM (full width at half maximum) of the  Mollow triplet peaks of the cavity-emitted spectrum ($S_c$) for various exciton-cavity detunings and exciton-cavity couplings as a function of the square of the Rabi frequency $\eta_x$. Clear signatures of phonon-induced EID are demonstrated by the  dependence of the FWHM on the square of the Rabi frequency; for smaller exciton-cavity couplings [Fig.~\ref{fig:fwhm1}(a) and Fig.~\ref{fig:fwhm1}(d)], this dependence is approximately linear (for the chosen exciton-cavity detuning). These two plots have opposite QD-cavity detunings, and EID is marginally larger when the cavity mode is detuned to the left of the QD due to increased phonon-induced coupling between the exciton and the cavity mode. This general trend of EID increasing  linearly with the square of the Rabi frequency  is in good agreement with recent experiments performed in semiconductor micropillars~\cite{stuttgart_prl}, where $g_{\rm exp}\approx16~\mu$eV [cf.~Fig.~\ref{fig:fwhm1}(d)]. Increasing the QD-cavity coupling $g$ also increases the FWHM of the sidebands due to stronger coupling between the QD and the cavity and enhanced emission into the cavity mode [Fig.~\ref{fig:fwhm1}(b) and Fig.~\ref{fig:fwhm1}(e)]. The dependence at higher QD-cavity couplings is nonlinear especially at low drives [see Figs.~\ref{fig:fwhm1}(c) and Fig.~\ref{fig:fwhm1}(f)];  we also find a linear increase in the FWHM as a function of the square of cw laser drive even in the absence of phonon coupling, primarily for large QD-cavity couplings. This effect can be attributed to enhanced emission into the cavity mode which is enhanced with increasing drive and broadens the triplet. We also observe a notable difference in the broadenings of the low-energy and high-energy components of the triplet in the presence of phonons. The triplet component proximal to the cavity mode shows larger broadening which increases with $g$.

In Fig.~\ref{fig:fwhm2} we study the numerically extracted FWHM of the off-resonance Mollow triplet as a function of the exciton-laser detuning $\Delta_{Lx}$. The coupling to electron-phonon scattering is seen to introduce additional broadening of the sidebands, as expected, due to EID; this can be seen from the differences in magnitudes between the plots in the top panel and the bottom panel. Also note that the two sidebands of the triplet are broadened by different amounts which is governed primarily by the proximity of the sideband to the cavity mode. In the absence of any phonon coupling, the Mollow component which is closer to the cavity mode is broadened more due to enhanced emission into the cavity. Moreover, the difference in the broadening of the high energy and the low energy components of the triplet increases with temperature. We also observe a noticeable difference in the broadening of the sidebands in the presence of phonons. We find an enhancement in the FWHM broadening when the cw laser drive is spectrally located between the cavity mode and the exciton mode. In contrast, there is substantial suppression of FWHM if the cw laser is detuned in the other direction relative to the exciton-cavity detuning. This suggest that phonons enhance the broadening process in the first scenario as it allows for an additional mechanism for enhanced emission into the cavity mode.

Recent experiments in semiconductor micropillars~\cite{stuttgart_prl} have noted a distinct spectral narrowing of both the high energy and the low energy sideband of the Mollow triplet with increasing exciton-cw laser detuning. This behaviour is quite contrary to what is obtained from a phonon-based EID model. However, a decrease in the Rabi frequency was also noticed, experimentally, as a function of detuning \cite{stuttgart_prl}. In addition,  more recent experimental data obtained for weakly coupled QD-cavity systems find the opposite dependence on the exciton-laser detuning wherein the FWHM of the off-resonance Mollow triplet is found to increase with the QD-laser detuning~\cite{stuttgart_private}.  Part of this anomaly can perhaps be attributed to additional detuning-dependent pure dephasing mechanisms~\cite{stuttgart_prl}. Further investigations and a closer collaboration between experiments and theory are likely required to understand these conflicting trends in more detail.

\section{Conclusions }
\label{conclusions}

We have presented a detailed theoretical analysis of the  fluorescence spectra  from a coherently driven cavity-QED system. Fundamentally distinctive features of the semiconductor QD system are found because of  acoustic phonon scattering  in the solid state environment, which we have modelled at a microscopic level.  The  nonperturbative treatment of the effects of QD exciton, acoustic phonons, cavity and the cw laser field on the coherent part of the Hamiltonian, was accomplished by a suitable choice of the decomposition of the system Hamiltonian. A polaron transformation was adopted that retains coherent phonon coupling to all orders, while introducing a suitable (modified) system-bath-interaction with which to perform perturbation theory that is accurate to second order. We derived a TC ME in the polaron frame to study the fluorescence spectra, and we showed that this ME produces identical numerical results to a NZ ME solution---the latter being significantly more difficult to solve numerically. We also showed that a Markovian ME  is perfectly valid for the cw pumping scenarios investigated in this paper, if one uses a polaron frame.

We then calculated the Mollow spectra as a function of exciton-cavity detuning
and found that electron-phonon-scattering causes EID and significant coupling between the exciton and cavity mode, throughout the entire phonon spectral function (easily spanning 5 meV). At a phonon bath temperature of $T=10~$K, the cavity resonance is found to go through a maximum at a specific
detuning, when the cavity is red shifted by about 1.4~meV. The resonance spectra
were  found to be strongly asymmetric with respect to exciton-cavity detuning. Next, we showed the role of pump excitation ($\eta_x$) and exciton-cavity coupling rate ($g$),
both of which are seen to have a significant impact
due to phonon-induced scattering processes.
Both the cavity-emitted spectra ($S_c$) and exciton-emitted spectra ($S_x$) were calculated
and these were found to have quite different spectral forms, especially
near the cavity mode resonance. We subsequently investigated the spectral FWHM of the Mollow triplet sidebands for various exciton-cavity detunings and exciton-cavity couplings. We  analysed the  Mollow triplet spectra as a function of the square of the Rabi frequency, and obtained unambiguous signatures of phonon-induced EID in accordance with recent experiments. We also studied the FWHM of the off-resonance Mollow triplet as a function of the exciton-laser detuning. The interaction of the QD with the phonon reservoir was found to introduce pronounced features in the fluorescence spectra exemplified by emission at the cavity mode and additional spectral broadening effects. With a systematic detuning between the driving laser and the exciton,  our theoretical results predict further  broadening which is a prediction that is in disagreement with recent experiments in semiconductor micropillars~\cite{stuttgart_prl}---which suggest a  spectral narrowing of sidebands with increasing exciton-laser detuning. Nevertheless, as mentioned above, recent experiments also show a spectral broadening with exciton--laser detuning~\cite{stuttgart_private}. 
The possible role of detuning-dependent pure dephasing also merits further study in this context, as relative magnitudes of pure dephasing and the QD radiative decay rate determine the curvature of the  FWHM plot for the off-resonant fluorescence sideband as a function of QD-laser detuning. Additional contributions from phonon-damped processes and higher order QD-phonon interactions may further influence the underlying physics of the laser-detuning-dependent EID. 
 

We  highlight that the semiconductor-specific features highlighted above are quite distinct
to the physics of atomic cavity-QED and in general the  application
of an atomiclike MEs to these systems may drastically  fail to produce the qualitative experimental features of semiconductor-based
cavity-QED systems. Although we have specialized our studies
to an exciton-driven system, it is straightforward to also
study a cavity-driven system as well, e.g., see Ref.~\onlinecite{roy_hughes_prx}.

\section*{Acknowledgments}
This work was supported by the National Sciences and Engineering Research Council of Canada. 
\\



\begin{thebibliography}{[1]}

\bibitem{yamamoto}Y. Yamamoto and A. Imamo\ifmmode \breve{g}\else \u{g}\fi{}lu, {\it Mesoscopic Quantum Optics}, Wiley-Interscience, 2001.

\bibitem{hohenester}U. Hohenester, {\it Optical properties of semiconductor nanostructures: Decoherence versus Quantum Control},
Handbook of Theoretical and Computational Nanotechnology (2006)

\bibitem{press}D. Press, S. G\"otzinger, S. Reitzenstein, C. Hofmann, A. L\"offler, M. Kamp, A. Forchel, and Y. Yamamoto,
Phys. Rev. Lett. {\bf98}, 117402 (2007).

\bibitem{entangled1} N. Akopian, N. H. Lindner, E. Poem, Y. Berlatzky, J. Avron, D. Gershoni, B. D. Gerardot, and P. M. Petroff,
Phys. Rev. Lett. {\bf 96}, 130501 (2006).

\bibitem{entangled2} A. Muller, W. Fang, J. Lawall and G. S. Solomon,
Phys. Rev. Lett. {\bf 103}, 217402 (2009).

\bibitem{entangled3} R. B. Patel, A. J. Bennett, K. Cooper, P. Atkinson, C. A. Nicoll, D. A. Ritchie and A. J. Shields,
Phys. Rev. Lett. {\bf 100}, 207405 (2008).


\bibitem{SC:StrongCoupling1}
see, for example,
J. P. Reithmaier, G. Seogonk, A. L\"offler, C. Hofmann, S. Kuhn, S. Reitzenstein,
L. V. Keldysh, V. D. Kulakovskii, T. L. Reinecke, and A. Forchel,
Nature  {\bf 432}, 197 (2004).

\bibitem{SC:StrongCoupling2}
T. Yoshie, A. Scherer, J. Hendrickson, G. Khitrova, H. M. Gibbs, G. Rupper, C. Ell, O. B. Shchekin, and D. G.
Deppe,
Nature \textbf{432}, 200 (2004).

\bibitem{SC:StrongCoupling3}
E. Peter, P. Senellart, D. Martrou, A. Lemaitre, J. Hours, J. M. G\'erard, and J. Bloch,
Phys. Rev. Lett.  {\bf 95}, 067401 (2005).

\bibitem{spinblockade}
A. Imamo\ifmmode \breve{g}\else \u{g}\fi{}lu, H. Schmidt, G. Woods, G. and M. Deutsch,
Phys. Rev. Lett. {\bf 79}, 1467 (1997).

\bibitem{atature}
A. N. Vamivakas, C. Y. Lu, C. Matthiesen, Y. Zhao, S. Falt, A. Badolato and M. Atature,
Nature {\bf 467}, 297 (2010).

\bibitem{muller}A. Muller, E. B. Flagg, P. Bianucci, X. Y. Wang, D. G. Deppe, W. Ma, J. Zhang, G. J. Salamo, M. Xiao, and C. K. Shih,
Phys. Rev. Lett. {\bf99}, 187402 (2007).

\bibitem{flagg}E. B. Flagg, A. Muller, J. W. Robertson, S. Founta, D. G. Deppe, M. Xiao, W. Ma, G. J. Salamo  and C. K. Shih,
Nature Physics {\bf 5}, 203 (2009).

\bibitem{vamivakas} A. N. Vamivakas, Y. Zhao, C.-Y. Lu and  M. Atature,
Nature Physics {\bf 5}, 198 (2009).

\bibitem{ates1}S. Ates, S. M. Ulrich, S. Reitzenstein, A. L\"offler, A. Forchel, A.  and P. Michler,
Phys. Rev. Lett. {\bf 103}, 167402 (2009).


\bibitem{ates2}S. Ates, S. M. Ulrich, A. Ulhaq, S. Reitzenstein, A. L\"offler, S. H\"ofling, A. Forchel  and P. Michler,
Nature Photonics {\bf3}, 724 (2009).


\bibitem{jelena1}A. Majumdar, A. Faraon, E. D. Kim, D. Englund, H. Kim, P. Petroff and J. Vu\ifmmode \check{c}\else \v{c}\fi{}kovi\ifmmode \acute{c}\else \'{c}\fi{},
Phys. Rev. B, {\bf 82}, 045306 (2010).


\bibitem{ates3}A. Ulhaq, S. Ates, S. Weiler, S. M. Ulrich, S. Reitzenstein, A. L\"offler, S. H\"ofling, L. Worschech, A. Forchel, and P. Michler,
Phys. Rev. B {\bf 82}, 045307 (2010).


\bibitem{stuttgart_prl}S. M. Ulrich, S. Ates, S. Reitzenstein, A. L\"offler, A. Forchel and P. Michler,
Phys. Rev. Lett., {\bf 106}, 247402 (2011).

\bibitem{ramsay1}A. J. Ramsay, T. M. Godden, S. J. Boyle, E. M. Gauger, A. Nazir, B. W. Lovett, A. M. Fox, and M. S. Skolnick,
Phys. Rev. Lett. {\bf105},177402 (2010).

\bibitem{ramsay2}A. J. Ramsay, Achanta Venu Gopal, E. M. Gauger, A. Nazir, B. W. Lovett, A. M. Fox, and M. S. Skolnick,
Phys. Rev. Lett. {\bf 104}, 017402 (2010).






\bibitem{besombes}L. Besombes, K. Kheng, L. Marsal and H. Mariette,
Phys. Rev. B {\bf 63}, 155307 (2001).



\bibitem{Favero:PRB03}
E. Peter, J. Hours, P. Senellart, A. Vasanelli, A. Cavanna, J. Bloch,
and J. M. G\'erard, Phys. Rev. B {\bf 69}, 041307 (2004).

\bibitem{Peter:PRB04}
I. Favero, G. Cassabois, R. Ferreira, D. Darson, C. Voisin, J. Tignon,
C. Delalande, G. Bastard, Ph. Roussignol, and J. M. G\'erard, Phys.
Rev. B {\bf 68}, 233301 (2003).






\bibitem{nazir2}D. P. S. McCutcheon and A. Nazir,
New J. Phys. {\bf 12}, 113042 (2010).

\bibitem{mogilevtsev1}D. Mogilevtsev, A. P. Nisovtsev, S. Kilin, S. B. Cavalcanti, H. S. Brandi, and L. E. Oliveira,
Phys. Rev. Lett. {\bf 100}, 017401 (2008).
\bibitem{mogilevtsev2}D Mogilevtsev, A P Nisovtsev, S Kilin, S B Cavalcanti, H S Brandi and L E Oliveira,
J. Phys.: Condens. Matter {\bf21}, 055801 (2009)


\bibitem{roy_hughes}C. Roy and S. Hughes,
Phys. Rev. Lett, {\bf 106}, 247403 (2011).


\bibitem{jelena_arka}A. Majumdar, E. D. Kim, Y. Gong, Yiyang, M. Bajcsy and J. Vu\ifmmode \check{c}\else \v{c}\fi{}kovi\ifmmode \acute{c}\else \'{c}\fi{}, 
Phys. Rev. B {\bf 84}, 085309 (2011).



\bibitem{HennessyNature:2007}
K. Hennessy, A. Badolato, M. Winger, A. Atature, S. Falt, E. L. Hu, and A. A.~Imamogl{\u u},
Nature {\bf 445}, 896 (2007).

\bibitem{KaniberPRB:2008}
M. Kaniber, A. Laucht, A. Neumann, J.M. Villas-Boas, M. Bichler, M.-C. Amann, and J. J. Finley,
Phys. Rev. B {\bf 77},
161303(R) (2008).


\bibitem{RuthOE:2007}
R. Oulton, B.D. Jones, S. Lam, A.R.A. Chalcraft, D. Szymanski, D. OBrien, T.F.
Krauss, D. Sanvitto, A. M. Fox, D.M. Whittaker, M. Hopkinson, and M.S. Skolnick,
  Opt. Express {\bf 15}, 17221 (2007).



\bibitem{SufczynskiPRL:2009}
J. Suffczynski, A. Dousse, K. Gauthron, A. Lemaitre, I. Sagnes, L. Lanco, J. Bloch, P. Voisin, and P. Senellart,
Phys. Rev. Lett.
{\bf 103}, 027401 (2009).

\bibitem{ota}Y. Ota, S. Iwamoto, N. Kumagai and Y. Arakawa,
arXiv:0908.0788 (2009).

\bibitem{TawaraOE:2009}
T. Tawara, H. Kamada, S. Hughes, H. Okamoto, M. Notomi, and T. Sogawa,
Opt. Express {\bf 17}, 6643 (2009).



\bibitem{Dalacu:PRB2010}
D. Dalacu, K. Mnaymneh, V. Sazonova, P. J. Poole, G. C. Aers, J. Lapointe, R. Cheriton,
A. J. SpringThorpe, and R. L. Williams,
%
Phys Rev B {\bf 82}, 033301 (2010).

\bibitem{Calic:PRL11}
M. Calic, P. Gallo, M. Felici, K. A. Atlasov, B. Dwir, A. Rudra, G. Biasiol, L. Sorba, G. Tarel, V. Savona, and E. Kapon
Phys. Rev. Lett. {\bf 106}, 227402 (2011).


\bibitem{JiaoJPC2008}
J. Xue, K-D Zhu and H. Zheng,
J. Phys. C {\bf 20}, 3252009 (2008).


\bibitem{HohenesterPRB:2009}
U. Hohenester, A. Laucht, M. Kaniber, N. Hauke,
A. Neumann, A. Mohtashami,
M. Selinger, M. Bichler,
and J. J. Finley, Phys. Rev. B {\bf 81}, 201311 (2009).

\bibitem{HohenesterPRB:2010}
U. Hohenester, Phys. Rev. B {\bf 81}, 155303 (2010).

\bibitem{kaer}P. Kaer, T. R. Nielsen, P. Lodahl, A.-P. Jauho, and J. M\o{}rk,
Phys. Rev. Lett. {\bf 104}, 157401 (2010).


\bibitem{hughes2} S. Hughes, P. Yao, F. Milde, A. Knorr, D. Dalacu, K. Mnaymneh, V. Sazonova, P. J. Poole, G. C. Aers, J. Lapointe, R. Cheriton  and R. L. Williams,
Phys. Rev. B {\bf 83}, 165313 (2011).

\bibitem{SavonaPRB:2010}
G. Tarel and V. Savona,
Phys. Rev. B {\bf 81}, 075305 (2010).


\bibitem{hughes1}  F. Milde, A. Knorr, and S. Hughes,
Phys. Rev. B {\bf 78}, 035330 (2008).

\bibitem{imamoglu}I. Wilson-Rae  and A. Imamo\ifmmode \breve{g}\else \u{g}\fi{}lu,
Phys. Rev. B {\bf65}, 235311 (2002)


\bibitem{me_book_1}H.-P. Breuer and F. Petruccione,
{\it The Theory of Open Quantum Systems}, Oxford University Press, 2002




\bibitem{mahan}G. D. Mahan, {\it Many-Particle Physics}, Plenum, New York, 1990Í

\bibitem{krum}B. Krummheuer, V. M. Axt, and T. Kuhn,
Phys. Rev. B {\bf 65}, 195313 (2002)


\bibitem{10}
P. Machnikowski, 
Phys. Rev. Lett. {\bf 96}, 140405 (2006).

\bibitem{11} E. A. Muljarov and R. Zimmermann,
 Phys. Rev. Lett. {\bf 93}, 237401 (2004).

\bibitem{12} G. Ortner, D. R . Yakovlev, M. Bayer, S. Rudin, T. L. Reinecke,
S. Fafard, Z . Wasilewski, and A. Forchel,
{\em Temperature dependence of the zero-phonon linewidth in InAsGaAs quantum dots},  Phys. Rev. B {\bf 70}, 201301(R) (2004).

\bibitem{13} S. Rudin, T. L. Reinecke, and M . Bayer,
 Phys. Rev. B {\bf 74}, 161305(R)
(2006).

\bibitem{14} G. Lindwall, A. Wacker, C . Weber, and A. Knorr,
Phys. Rev. Lett. {\bf 99}, 087401 (2007).

\bibitem{carmichael}H. J. Carmichael and D. F. Walls,
J. Phys. A: Math. Nucl. Gen. {\bf 6}, 1552 (1973).


\bibitem{roy_hughes_prx} C. Roy and S. Hughes,
Phys. Rev. X {\bf 1}, 021009 (2011).



\bibitem{me_1}H.-P. Breuer, B. Kappler and F. Petruccione,
Phys. Rev. A {\bf59}, 1633 (1999).

\bibitem{me_2}J. Fischer and H.-P. Breuer,
Phys. Rev. A {\bf 76} 052119 (2007).

\bibitem{me_3}A. Smirne and B. Vacchini,
Phys. Rev. A, {\bf 82}, 022110 (2010).

\bibitem{me_4}H. J. Carmichael, {\it Statistical Methods in Quantum Optics 1: Master Equations and Fokker-Planck Equations},
Springer,  2003.

\bibitem{zubairy}M. Scully and M. Zubairy, {\it Quantum Optics}, Cambridge University Press (1997).




\bibitem{quadratic_phonon} E. A. Muljarov and R. Zimmermann,
 Phys. Rev. Lett. {\bf 93}, 237401 (2004).

\bibitem{Rudin:PRL06}
S. Rudin, T. L. Reinecke, and M. Bayer, Phys. Rev. B {\bf 74}, 161305(R)
(2006).

\bibitem{BorriPRL:2001} P. Borri, W. Langbein, S. Schneider, U. Woggon, R. L. Sellin, D.
Ouyang, and D. Bimberg, Phys. Rev. Lett. {\bf 87}, 157401 (2001).

\bibitem{roy}C. Roy  and S. John,
Phys. Rev. A {\bf 81}, 023817 (2010)







\bibitem{nazir1}A. Nazir,
Phys. Rev. B {\bf 78}, 153309 (2008).

\bibitem{nazir_new}D. P. S. McCutcheon, N. S. Dattani, E. M. Gauger, B. W. Lovett and A. Nazir,
Phys. Rev. B, {\bf 84}, 081305 (2011).


\bibitem{axt_new}
M. Gl\"assl, A. Vagov, S. L\"uker, D. E. Reiter, M. D. Croitoru, and P. Machnikowski, V. M. Axt, and T. Kuhn
Phys. Rev. B {\bf 84}, 195311 (2011).

\bibitem{QOToolbox}
S. M. Tan 1999 J. Opt. B: Quantum Semiclass. Opt. {\bf 1}, 424 (1999).


\bibitem{stuttgart_private} A. Ulhaq, S. Ulrich, and P. Michler, {\it Private communication}.





\end{thebibliography}
\end{document}